\documentclass[a4paper]{article}

\usepackage{enumerate}
\usepackage{latexsym}
\usepackage{citesort}
\usepackage{amsmath,amssymb,amsfonts}
\usepackage{graphicx}
\usepackage{textcomp}
\usepackage{textcomp}
\usepackage{caption}
\usepackage{subcaption}
\usepackage{wrapfig}
\usepackage{booktabs}
\usepackage{comment}
\usepackage{amssymb}
\usepackage{setspace}

\usepackage{algorithm,algorithmic} 

\newtheorem{thm}{Theorem}

\newcommand{\Env}{\mathcal{Z}}
\newcommand{\Sim}{\mathcal{S}}
\newcommand{\Adv}{\mathcal{A}}

\newcommand{\Fcr}{\mathcal{F}^{*}_{\mathrm{CR}}}
\newcommand{\Fml}{\mathcal{F}^{*}_{\mathrm{ML}}}
\newcommand{\Ff}{\mathcal{F}^{*}_{f}}
\newcommand{\mFf}{\mathcal{F}^{*}_{f,\mathrm{neq}}}
\newcommand{\phisr}{\phi_{s,r}}
\newcommand{\phirf}{\phi_{\mathrm{refund}}}

\newcommand{\coins}{{\sf coins}}

\newcommand{\SW}{{\sf SimWallet}}

\newcommand{\Fcrg}{\mathcal{F}^{*}_{\mathrm{CRG}}}

\newtheorem{defi}{Definition}
\title{Constant-Round Linear-Broadcast \\ Secure Computation with Penalties
}

\author{Takeshi Nakai$^{1}$, Kazumasa Shinagawa$^{2}$ $^{3}$   \\
        \small $^{1}$ Toyohashi University of Technology, Aichi, Japan  \\
        \small $^{2}$ Ibaraki University, Ibaraki, Japan \\
        \small $^{3}$ National Institute of Advanced Industrial Science and Technology, Tokyo, Japan
}

\begin{document}
\maketitle



\begin{abstract}
It is known that Bitcoin enables achieving fairness in secure computation by imposing monetary penalties on adversarial parties. 
This functionality is called \textit{secure computation with penalties}.
Bentov and Kumaresan (Crypto 2014) introduced the \textit{claim-or-refund} functionality that can be implemented via Bitcoin. They achieved secure computation with penalties with $O(n)$ rounds and $O(n)$ broadcasts for any function, where $n$ is the number of parties.
After that, Kumaresan and Bentov (CCS 2014) showed a constant-round protocol. Unfortunately, this protocol requires $O(n^2)$ broadcasts.
As far as we know, no protocol achieves $O(1)$ rounds and $O(n)$ broadcasts based on Bitcoin.
This work accomplishes such efficiency in secure computation with penalties.
We first show a protocol in a slightly relaxed setting called \textit{secure computation with non-equivalent penalties}.
This setting is the same as secure computation with penalties except that every honest party receives \textit{more than} a predetermined amount of compensation, while the previous one requires that every honest party receives the same amount of compensation. Namely, our setting allows the compensations for honest parties to be non-equivalent.
Moreover, we present a technique to remove the non-equivalence of our protocol without sacrificing efficiency. We then propose a new ideal functionality called \textit{claim-refund-or-give} that can be implemented via Bitcoin.
\end{abstract}


\section{Introduction} \label{Sect:Intro}
\subsection{Backgrounds}
Secure computation enables distrustful parties to jointly compute a function on their private inputs \cite{Yao86}.
There are several notions of security, such as privacy, correctness, independence of inputs, guaranteed output delivery, and fairness. 
Fairness requires that at the end of a protocol, either all parties learn the output value or none of them learn it.
Namely, it guarantees that no corrupted parties aborts the protocol, with only the aborting parties learning the output value.
Unfortunately, it is known that fairness cannot always be achieved in the standard model if a majority of parties are corrupted \cite{Cleve86}.

There are works to circumvent the impossibility result by imposing monetary penalties on aborting parties \cite{Lindell08}.
It prevents the aborting attack since a corrupted party loses money when he/she aborts the protocol. The monetary penalty mechanism can be implemented by decentralized cryptocurrencies, e.g., Bitcoin \cite{Nakamoto09}. 
There is a line of research on constructing cryptographic protocols using decentralized cryptocurrencies \cite{Back14,MarcinSP14,MarcinFC14,Kumaresan14,Bentov14,Kumaresan15,Zhao15,Ranjit16,Kumaresan16,Bartoletti17,Miller17}.

Back and Bentov \cite{Back14} and Andrychowicz, Dziembowski, Malinowski, and Mazurek \cite{MarcinSP14} introduced secure computation on Bitcoin.
They studied fair lottery protocols that guarantee any aborting party learning the output is forced to pay penalties to all other parties.
After that, Bentov and Kumaresan \cite{Bentov14} formalized such a model of computation as \textit{secure computation with penalties}.
In particular, they defined a \textit{claim-or-refund functionality} $\Fcr$ that plays a crucial role in secure computation with penalties.
It enables a sender to send coins with a predicate $\phisr(\cdot)$ and a time period $\tau$ to a receiver. 
The receiver gets the coins if he/she reveals a witness $w$ such as $\phisr(w)=1$ in $\tau$, and the sender gets back the coins if the receiver does not publish the solution in $\tau$. 
They showed that secure computation with penalties
can be realized for any function in the $(\mathcal{F}_{\mathrm{OT}}, \Fcr)$-hybrid model, where
$\mathcal{F}_{\mathrm{OT}}$ is an ideal functionality of oblivious transfer.
Their protocol requires $O(n)$ rounds and $O(n)$ broadcasts, where 
$n$ is the number of parties. Here, the number of broadcasts refers to the number of transactions in Bitcoin.
Kumaresan and Bentov \cite{Kumaresan14} introduced an ideal transaction functionality
$\Fml$ and showed that secure computation with penalties can be realized in 
the $(\mathcal{F}_{\mathrm{OT}}, \Fml)$-hybrid model with only $O(1)$ rounds.\footnote{Roughly speaking, $\Fml$ allows multiple parties to lock their coins. Each party gets the coins back only if he/she reveals a predetermined private witness.}
However, this protocol requires $O(n^2)$ broadcasts. 
To the best of our knowledge, there is no protocol for secure computation with
penalties on Bitcoin with $O(1)$ rounds and $O(n)$ broadcasts.

\subsection{Related works and the scope of this paper}
Kumaresan, Moran, and Bentov \cite{Kumaresan15} extended 
secure computation with penalties to the \textit{reactive model} 
that can also handle multistage functionalities, such as Texas Holdem Poker. 
Also, they defined a new security model for the reactive model and proposed
a fair protocol in the reactive model.
This paper focuses on the single-stage (i.e., non-reactive) model following the same setting in \cite{Bentov14}. 

Kumaresan and Bentov \cite{Ranjit16} improved the efficiency of protocols by amortizing the cost over multiple executions. 
Kumaresan, Vaikuntanathan, and Vasudevan \cite{Kumaresan16} reduced the script complexity of $\phisr$. 
We focus on reducing the number of rounds and the number of broadcasts.

There is an approach to combining covert security and monetary penalty. 
It provides a mechanism for imposing financial penalties on parties whose malicious behavior is detected \cite{Zhu19,Faust22}. 
This paper focuses on the aborting attack and does not deal with other malicious behaviors.

To realize monetary penalties, every party must pay deposit money at the beginning of a protocol. 
There are works to reduce the deposit amount \cite{Bartoletti17,Miller17} since a large deposit amount may discourage parties from participating in the protocol.
In this paper, we also tackle to reduce the deposit amount. 

There are several works \cite{Kiayias16,Bentov17,David18,Miller17,David19} based on the model with stateful contracts, e.g., Ethereum \cite{Wood14}. 
This model is stronger than our model since it requires an advanced blockchain techniques beyond Bitcoin. 

\subsection{Our contributions}
We introduce new functionality, \textit{secure computation with non-equivalent penalties}, which is a slightly relaxed model of secure computation with penalties. 
It guarantees that each honest party is compensated with \textit{more than} a predetermined amount of coins, while secure computation with penalties guarantees that every honest party is compensated for the \textit{same} amount of coins. 
That is, two honest parties may be compensated with different amounts of coins, although they are at least a predetermined amount. 
We note that the non-equivalent penalty also prevents the aborting attack since it is the same with secure computation with penalties in that it imposes a monetary penalty on the aborting parties.

See Table \ref{Tbl:Result}, which summarizes our contributions.
We first show that secure computation with non-equivalent penalties can be realized in the $(\mathcal{F}_{\mathrm{OT}}, \Fcr)$-hybrid model. 
Our technical contribution is to propose a new \textit{fair reconstruction protocol}, which is a subprotocol of a secure computation protocol with (non-equivalent) penalties, with $O(1)$ rounds and $O(n)$ broadcasts. 
As a result, we obtain a secure computation protocol with non-equivalent penalties with $O(1)$ rounds and $O(n)$ broadcasts by using our fair reconstruction protocol in the Bentov--Kumaresan's construction. 

Although this protocol achieves a constant round, there are two issues to be addressed. 
The first issue is to increase the deposit amount. While the deposit amount of Bentov--Kumaresan's protocol is $O(nq)$, our protocol requires $O(n^2q)$, where $q$ is a parameter of the penalty amount. To solve the first issue, we propose a technique to reduce the deposit by about $1/(l+1)$ increasing the number of rounds by $2l$. 

The second issue is the non-equivalence of compensations. While our protocol achieves fairness, it is undesirable to force all parties to accept the \textit{unfairness} of the compensations.
To solve the second issue, we propose a new ideal functionality called \textit{claim-refund-or-give functionality}, which is an extension of $\Fcr$, and show that it can be used to construct a secure computation protocol with penalties, with $O(1)$ rounds and $O(n)$ broadcasts. 

\subsection{Organization}
The remaining part of this paper is organized as follows:
In Section \ref{Sect:Prel}, 
we introduce basic notations, a security model, and  ideal functionalities.
Section \ref{Sect:SCP} describes Bentov--Kumaresan's
protocol for secure computation with penalties \cite{Bentov14}.
Section \ref{Sect:PP} introduces secure computation with non-equivalent penalties and proposes a protocol for it with 
$O(1)$ rounds and $O(n)$ broadcasts. 
Section \ref{Sect:Reduce} shows a technique to reduce the deposit
amount of our protocol.
Section \ref{Sect:Equiv} presents a technique for removing
the non-equivalence of compensation amounts from our protocol.
We conclude this paper in Section \ref{Sect:CR}.
(Sections \ref{Sect:Reduce}--\ref{Sect:Equiv} 
 and the appendixes are newly added from the earlier version \cite{Nakai21}.)

\begin{table}[t]
\centering
\caption{Comparison of secure computation protocols with penalties: $n$ is the number of parties, and $q$ refers to a parameter of the penalty amount.}
  \begin{tabular}{ccccc} 
    \toprule
    ~  \,& $\#$ of  \,& $\#$ of \,& Deposit \,&  \\
    ~References  \,& Rounds \,& Broadcasts \,& amount \,& Compensations \\
    \midrule \midrule
    Bentov--Kumaresan \cite{Bentov14} & $O(n)$ & $O(n)$ & $O(nq)$ & Equivalent     \\
    \midrule
    Kumaresan-Bentov \cite{Kumaresan14} & $O(1)$ & $O(n^2)$ & $O(nq)$  & Equivalent     \\
    \midrule
    This work (Sect. \ref{Sect:PP})& $O(1)$ & $O(n)$  & $O(n^2q)$ & Non-equivalent \\
    \midrule
    This work (Sect. \ref{Sect:Equiv})& $O(1)$ & $O(n)$  & $O(n^2q)$ & Equivalent \\
    \bottomrule
  \end{tabular}
    \label{Tbl:Result}
\end{table}


\section{Preliminaries} \label{Sect:Prel}
\subsection{Basic notations}
For any positive integer $i \in \mathbb{N}$, $[i]$ denotes the set of integers $\{1,\ldots,i\}$.
We denote by $n$ the number of parties in a protocol.
We denote by $H \subseteq [n]$ (resp. $C \subseteq [n]$) the set of honest (resp. corrupted) parties. 
Since each party is either honest or corrupted, it must hold $h+c = n$ for $h := |H|$ and $c := |C|$.
We consider settings where $c < n$.
We denote by $k$ a security parameter. 
We assume that all parties are non-uniform probabilistic polynomial-time algorithms in $k$.

\subsection{Secure computation with coins}
Bentov--Kumaresan \cite{Bentov14} introduced a new secure computation model called \textit{secure computation with coins} (SCC) model. 
It is the same model as the standard model except that entities (i.e., parties, adversaries, ideal functionalities, and an environment) can deal with a non-standard entity called \textit{coins}, which is an atomic entity representing electronic money. 
Coins are assumed to be having the following properties.
\begin{itemize}
 \item Coins cannot be duplicated and forged. 
 \item No multiple parties hold the same coin simultaneously.
 \item Any parties can transfer their coins to other parties freely.
 \item Each coin is perfectly indistinguishable from one another.
\end{itemize}
We use the notation $\coins(\cdot)$ to express the amount of coins.
If a party owning $\coins(x)$ receives $\coins(y)$ from another party,
then the party holds $\coins(x+y)$ as a result.

In the SCC model, some ideal functionalities can deal with coins. 
We call such a functionality a \textit{special ideal functionality}. 
These functionalities are described with the superscript $*$, e.g.,
$\mathcal{F}^{*}_{\mathrm{xxx}}$.
We call an ideal functionality without handling coins a \textit{standard} ideal functionality. 
Our protocol is realized in the hybrid model where parties have access to
a standard functionality $\mathcal{F}_{\mathrm{OT}}$, which is the
ideal functionality for oblivious transfer, and a special ideal functionality $\Fcr$, described later.

The SCC model follows the real/ideal simulation paradigm as with the standard secure computation model. 
Let $\mathrm{IDEAL}_{\mathcal{F},\Sim, \Env}(k,z)$ denote the output of an environment $\Env$ in the ideal world for realizing an ideal functionality $\mathcal{F}$, where $\Env$ (with an auxiliary input $z$) is interacting with an ideal adversary $\Sim$ on security parameter $k$. 
Let $\mathrm{HYBRID}^{\mathcal{G}}_{\pi, \Adv, \Env}(k,z)$ denote the output of environment $\Env$ in the real (hybrid) world for executing a hybrid protocol $\pi$ with an ideal functionality $\mathcal{G}$, where $\Env$ is interacting with a real adversary $\Adv$. 
The difference with the standard secure computation is that all entities (i.e., parties, adversaries, special ideal functionalities, and an environment) can deal with coins: sending coins, storing coins, and receiving coins. 

\begin{defi}
\label{Def:SCC}
Let $\pi$ be a probabilistic polynomial-time protocol with $n$ parties, and $\mathcal{F}$ be a probabilistic polynomial-time $n$-party (standard or special)
ideal functionality. We say that $\pi$ {\sf SCC} realizes $\mathcal{F}$
with abort in the $\mathcal{G}$ hybrid model (where $\mathcal{G}$ is a
standard or special ideal functionality) if for every non-uniform 
probabilistic polynomial-time adversary $\Adv$, there exists a non-uniform
probabilistic polynomial-time adversary $\Sim$ such that for every non-uniform probabilistic polynomial-time environment $\Env$, two families of probability distributions
$\{\mathrm{IDEAL}_{\mathcal{F},\Sim, \Env}(k,z)\}_{k\in\mathbb{N},z\in\{0,1\}^*}$
and
$\{\mathrm{HYBRID}^{\mathcal{G}}_{\pi, \Adv, \Env}(k,z)\}_{k\in\mathbb{N},z\in\{0,1\}^*}$
are computationally indistinguishable.
\end{defi}

\subsection{Special ideal functionalities}\label{Subsect:func}
\subsubsection{Claim-or-refund functionality $\Fcr$}
This functionality $\Fcr$ \cite{Bentov14} can be seen as an analogue of puzzles with bounty. 
Roughly speaking, for a puzzle $\phisr$ with coins submitted by a sender, a receiver gets the coins if and only if he/she submits a solution $w$ of the puzzle (i.e., $\phisr(w) = 1$). 
$\Fcr$ consists of three phases: deposit, claim, and refund. 
In the deposit phase, a sender $P_s$ sends to a receiver $P_r$ ``conditional'' coins together with a circuit $\phisr$. 
The coins also have a round number $\tau$ specified by the sender.
In the claim phase, the receiver $P_r$ claims to receive the coins. 
$P_r$ can receive the coins only if he/she broadcasts the witness $w$ of $\phisr$ (i.e., $\phisr(w) = 1$) in $\tau$.
Note that the witness $w$ published in the claim phase is made public to all parties.
In the refund phase, if $P_r$ does not claim in $\tau$, then the coins are refunded to the sender $P_s$. 
See Functionality \ref{Alg:Fcr} for a formal description of $\Fcr$.\footnote{\cite{Bentov14ep,Ranjit16} show how to realize $\Fcr$ using Bitcoin.}
At least one broadcast is necessary to realize $\Fcr$ on Bitcoin.
Thus, the number of calling $\Fcr$ corresponds to the number of broadcasts.

We call the message in the deposit phase a deposit transaction. 
We use the following ``arrow'' notation to denote the deposit transaction for the sender $P_s$ and the receiver $P_r$. 
\begin{align*}
 P_s \xrightarrow[\,\,\,\,\,\,\,\,\,c, \tau\,\,\,\,\,\,\,\,\,]{w} P_r
\end{align*}
After making an arrow from $P_r$ to $P_s$ as above (i.e., after the deposit phase), $P_r$ can claim to receive $\coins(c)$ only if he/she publishes the witness $w$
in round $\tau$. 
$\coins(c)$ is refunded back to the original holder $P_s$ if $P_r$ does not publish $w$ in $\tau$.

\subsubsection{Secure computation with penalties $\Ff$}
This functionality $\Ff$ is the same as the standard secure function evaluation except that aborting players are forced to pay penalties \cite{Bentov14}. 
In principle, it guarantees the following properties.
\begin{itemize}
 \item No honest party pays any penalty.
 \item If a party aborts after learning the output value and does not tell 
       the value to the other parties, then every party who does not learn the value
       is compensated with coins.
\end{itemize}

See Functionality \ref{Alg:Ff} 
for a formal description of $\Ff$. 
The parameters $q$ and $d$ specify the amounts of coins. 
At the beginning of the protocol, each party submits $\coins(d)$ together with input $x_i$. 
If a party aborts after learning the output and does not tell the value to the other parties, then $\Ff$ gives $\coins(q)$ to every party who does not learn the output as compensation. 
Then, it is important note that the compensation amount is always $q$ for any parties.

$H$ is a set of honest parties and $H' \subseteq H$ is a subset chosen by $\Sim$, which represents parties who are compensated. 
At first glance, it is somewhat strange that $\Sim$ chooses a subset of honest parties. 
The reason why $H'$ is needed is that there are two types of aborting in secure computation with abort. 
The first one is that an adversary aborts after obtaining the output and thus honest parties cannot obtain the outputs. 
In this case, $\Sim$ chooses $H' = H$ and all honest parties are compensated with $\coins(q)$ although the output is stolen by the adversary. 
The second one is that an adversary aborts before obtaining the output so the protocol just terminates. 
In this case, $\Sim$ chooses $H' \subsetneq H$ 
(possibly empty) 
and the parties in $H'$ are compensated with $\coins(q)$.\footnote{$H''$ is required for a technical reason in order to prove the security. See \cite{Bentov14} for a detail. In order to prove the security of our protocol, our new functionality follows the same strategy.}

\subsection{Non-malleable secret sharing with public verifiability and public reconstructibility}\label{Sect:pubNMSS}
A \textit{non-malleable secret sharing scheme with public verifiability and public reconstructibility} (in short, {\sf pubNMSS}) \cite{Bentov14} is a variant of non-malleable secret sharing scheme. 
The share algorithm of {\sf pubNMSS} takes a secret $s$ as input,
generates ``tag-token'' pairs $({\sf Tag}_i, {\sf Token}_i)_{i\in[n]}$, and
outputs ${\sf Token}_i$ and $({\sf Tag}_1,\dots, {\sf Tag}_n)$ to each party $P_i$.
The parties can reconstruct $s$ by collecting all $t$ tokens,
where $t$ is a predetermined threshold value.
For all $i \in [n]$, the parties can verify if the published ${\sf Token}_i$ 
is valid with ${\sf Tag}_i$.
The tag-token pairs have the following properties.
\begin{itemize}
 \item All tags $({\sf Tag}_1,\dots, {\sf Tag}_n)$ leak no information about $s$.
 \item Any sets of $t' (< t)$ tokens leak no information about $s$.
 \item For any $i \in [n]$, the adversary cannot generate 
       ${\sf Token}'_i (\neq {\sf Token}_i)$ such that $({\sf Tag}_i, {\sf Token}'_i)$
       is a valid tag-token pair.
\end{itemize}

A {\sf pubNMSS} scheme can be obtained from the \textit{honest-binding commitment}, 
which can be constructed from one-way functions \cite{Garay11}.
Roughly speaking, the honest binding commitment is a commitment that a malicious sender can decommit to any value, which is known as the \textit{equivocation} property.

${\sf Tag}_i$ is an (honest-binding) commitment that is computed by 
a secret share $sh_i$ and a randomness $r_i$ as input, and
${\sf Token}_i := (sh_i, r_i)$. 
Namely, the parties can verify if the published ${\sf Token}'_i = (sh'_i, r'_i)$ is 
valid by comparing ${\sf Tag}_i$ and the commitment whose input is $sh'_i$ and $r'_i$.
In the following discussions, this verification corresponds to $\phisr$
in $\Fcr$ executions.

\begin{algorithm}[htbp]
\floatname{algorithm}{Functionality}
\caption{Claim-or-refund functionality  $\Fcr$}
\label{Alg:Fcr}
 \begin{description}
 \item[\textbf{Setup}:]  The session identifier is $sid$. 
              Running with parties $P_1, \dots, P_n$ and an ideal adversary $\mathcal{S}$.
 \item[\textbf{Deposit phase}:] 
 Receiving $(\mathrm{deposit}, sid, ssid, s, r, \phisr, \tau, \coins(c))$ from $P_s$,
 perform the following process.
\begin{enumerate}
 \item Record the message $(\mathrm{deposit}, sid, ssid, s, r, \phisr, \tau, c)$
 \item Send all parties $(\mathrm{deposit}, sid, ssid, s, r, \phisr, \tau, c)$
 \renewcommand{\labelitemi}{-}
 \begin{itemize}
  \item Ignore any future messages with the same $ssid$ from $P_s$ to $P_r$.
 \end{itemize}
\end{enumerate}
 \item[\textbf{Claim phase}:]
 Receiving $(\mathrm{claim}, sid, ssid, s, r, \phisr, \tau, c, w)$ 
 from $P_r$ in round $\tau$, perform the following process.
 \begin{enumerate}
  \item Check the two conditions:
 \renewcommand{\labelitemi}{-}
  \begin{itemize}
     \item $(\mathrm{deposit}, sid, ssid, s, r, \phisr, \tau, c)$ was recorded,
     \item $\phisr(w) = 1$.
  \end{itemize}
  \item If both checks are passed, perform the following process:
	\begin{enumerate}
	\setlength{\leftskip}{0.5cm}
	 \item send $(\mathrm{claim}, sid, ssid, s, r, \phisr, \tau, c, w)$ to all parties,
	 \item send $(\mathrm{claim}, sid, ssid, s, r, \phisr, \tau, \coins(c))$ to $P_r$,
	 \item delete the record $(\mathrm{deposit}, sid, ssid, s, r, \phisr, \tau, c)$.
	\end{enumerate}
\end{enumerate}
 \item[\textbf{Refund phase}:] 
 In $\tau + 1$, if the record $(\mathrm{deposit}, sid, ssid, s, r, \phisr, \tau, c)$ was not deleted, then perform the following process:
 \begin{enumerate}
  \setlength{\leftskip}{0.5cm}
  \item send $(\mathrm{refund}, sid, ssid, s, r, \phisr, \tau, \coins(c))$ to $P_s$,
  \item delete the record $(\mathrm{deposit}, sid, ssid, s, r, \phisr, \tau, c)$.
 \end{enumerate}
 \end{description}
\end{algorithm}

\begin{algorithm}[htbp]
\floatname{algorithm}{Functionality}
 \caption{Secure computation with penalties $\Ff$}
 \label{Alg:Ff}
 \begin{description}
  \item[\textbf{Setup}:] 
  The session identifier is $sid$. 
  Running with parties $P_1, \dots, P_n$, 
  and an ideal adversary $\mathcal{S}$ that corrupts parties $\{P_i\}_{i\in C}$. 
  Let $d$ be a parameter representing the safety deposit, and let $q$ denote 
  the penalty amount.
 \item[\textbf{Input phase}:] 
 Wait to receive the following messages.
   \renewcommand{\labelitemi}{-}
 \begin{itemize}
  \item  $(\mathrm{input}, sid, ssid, i, x_i, \coins(d))$ from $P_i$ for all $i \in H$
  \item  $(\mathrm{input}, sid, ssid, \{y_i\}_{i\in C}, H', \coins(h'q))$ 
         from $\mathcal{S}$, where $H' \subseteq H$ and $h' = |H'|$
 \end{itemize}
 \item[\textbf{Output phase}:] 
 Perform the following process.
 \begin{enumerate}
  \item Send $(\mathrm{return}, sid, ssid, \coins(d))$ to each $P_i$ for $i \in H$.
  \item Compute $(y_1, \dots, y_n) \leftarrow f(x_1, \dots, x_n)$.
  \renewcommand{\labelitemi}{-}
  \begin{itemize}
   \item if $h' =0$, then send message $(\mathrm{output}, sid, ssid, y_i)$ to $P_i$ for 
         $i \in H$, and terminate.
   \item If $0< h' <h$, then send $(\mathrm{extra}, sid, ssid, \coins(q))$ to $P_i$
         for each $i \in H'$, and terminate, where $h:=|H|$.
   \item If $h' = h$, then send message $(\mathrm{output}, sid, ssid, \{y_i\}_{i \in C})$
         to $\mathcal{S}$.
  \end{itemize}
  \item If $\mathcal{S}$ returns $(\mathrm{continue}, sid, ssid, H'')$, 
	where $H'' \subseteq H$,
	then perform the following process:
	\begin{enumerate}
	 \setlength{\leftskip}{0.5cm}
	 \item send $(\mathrm{output}, sid, ssid, y_i)$ to $P_i$ 
               for all $i \in H$,
	 \item send $(\mathrm{payback}, sid, ssid, \coins((h-h'')q))$ to $\mathcal{S}$ 
	       where $h'' = |H''|$,
	 \item send $(\mathrm{extrapay}, sid, ssid, \coins(q))$ to $P_i$ for each 
	       $i \in H''$.
	\end{enumerate}
  \item Else if $\mathcal{S}$ returns $(\mathrm{abort}, sid, ssid)$, send
	$(\mathrm{penalty}, sid, ssid, \coins(q))$ to $P_i$ for all $i \in H$.
 \end{enumerate}
 \end{description}
\end{algorithm}

\section{Existing Protocol for Secure Computation with Penalties} \label{Sect:SCP}
In this section, we introduce Bentov--Kumaresan's protocol \cite{Bentov14} 
for secure computation with penalties in the $(\mathcal{F}_{\mathrm{OT}}, \Fcr)$-hybrid model.

\subsection{Bentov--Kumaresan's protocol} \label{Sect:OV}

For a function $f$, an \textit{augmented function} denoted by $\hat f$ is defined by a function that takes an input $x$ and distributes secret shares of the output value $f(x)$. 
The underlying secret sharing scheme is non-malleable secret sharing with publicly verifiability and publicly reconstructibility (Section \ref{Sect:pubNMSS}), where the threshold value is $n$. 
Thus the augmented function $\hat f$ outputs a token ${\sf Token}_i$ (i.e., a share of $f(x)$) and a set of tags $({\sf Tag}_1,\dots, {\sf Tag}_n)$ to party $P_i$. 

Bentov--Kumaresan's protocol proceeds as follows: 
\begin{enumerate}[(i)]
\item The parties execute a secure computation protocol for $\hat f$, and then each party $P_i$ obtains a token ${\sf Token}_i$ of $f(x)$ and a set of tags $({\sf Tag}_1,\dots, {\sf Tag}_n)$. (Note that this is the standard computation without Bitcoin). 
\item For the reconstruction of tokens, the parties execute the \textit{fair reconstruction protocol}, where each party $P_i$ is forced to broadcast a token ${\sf Token}_i$. The validity of the submitted token ${\sf Token}_i$ is verified with the tag ${\sf Tag}_i$. 
(Note that this computation is based on Bitcoin).
\end{enumerate}

It is well known that the OT functionality $\mathcal{F}_{\mathrm{OT}}$ is sufficient to achieve secure computation for any standard functionality \cite{Kilian88,Ishai08}.
Moreover, this can be performed in constant rounds \cite{Ishai08}.
Therefore, the secure computation stage (i) is performed in constant rounds in the $\mathcal{F}_{\mathrm{OT}}$-hybrid model. 

The main step of Bentov--Kumaresan's protocol is the fair reconstruction protocol (ii). 
By collecting all tokens, the parties can reconstruct the output value $f(x)$. 
However, malicious parties may abort so as to learn the output value while other parties do not. 
The fair reconstruction protocol prevents parties from aborting in the reconstruction phase. 
When malicious parties abort, they have to pay some amount of money for compensation to honest parties. 
It satisfies the following conditions:
\begin{enumerate}[(A)]
 \item No honest party pays any penalty.
 \item If an adversary learns the reconstruction result, but an honest party cannot,
       then the honest party is compensated with coins. 
       Furthermore, the compensation amounts are the same for any honest parties.
\end{enumerate}
Note that honest parties are not guaranteed to receive compensation if an adversary aborts without learning the output value.

In summary, secure computation with penalties can be realized by executing a secure computation protocol for $\hat f$ and the fair reconstruction protocol. 
The next section shows Bentov--Kumaresan's fair reconstruction protocol in the $\Fcr$-hybrid model. 

\subsection{Bentov--Kumaresan's fair reconstruction protocol} \label{Sect:FRP}

Hereafter, we use $T_i$ to denote ${\sf Token}_i$.
Suppose that each party $P_i$ has a token $T_i$ and a set of tags $({\sf Tag}_1, \dots, {\sf Tag}_n)$ at the beginning of the fair reconstruction protocol.
We assume that all parties agree on the penalty amount $q$, where honest parties are compensated with $\coins(q)$ when malicious parties abort with obtaining the output value. 
In the below, we successively explain a na\"{\i}ve approach, a solution for the two-party setting, and a solution for the $n$-party setting. 

\subsubsection*{Na\"{\i}ve approach}
Suppose that the number of parties is two. 
A na\"{\i}ve approach is to make a deposit transaction from $P_1$ to $P_2$ and a deposit transaction of the reverse direction as follows:
\begin{align}
 P_1 \xrightarrow
     [\,\,\,\,\,\,\,\,\,q, \tau\,\,\,\,\,\,\,\,\,]{T_2} P_2  \tag{1}\label{Eq:2-naive1} \\
 P_2 \xrightarrow
     [\,\,\,\,\,\,\,\,\,q, \tau\,\,\,\,\,\,\,\,\,]{T_1} P_1  \tag{2}\label{Eq:2-naive2} 
\end{align}
The above arrow means that ``$P_2$ can receive $\coins(q)$ only if $P_2$ publishes the token $T_2$, otherwise $\coins(q)$ is refunded back to $P_1$'' (see Section \ref{Subsect:func}). 
The bottom arrow is similar. 
At first glance, it seems a fair reconstruction protocol satisfying conditions (A) and (B) in Section \ref{Sect:OV}. 
However, it is not the case. 
For instance, when $P_2$ is malicious, $P_2$ can steal $\coins(q)$ from $P_1$ as follows:
after establishing transaction \eqref{Eq:2-naive1}, $P_2$ publishes the token $T_2$ without making transaction \eqref{Eq:2-naive2}.
As a result, honest $P_1$ loses $\coins(q)$. 
This violates condition (A).

\subsubsection*{Bentov-Kumaresan's solution}
In order to avoid the above attack, Bentov--Kumaresan's fair reconstruction protocol for the two-party setting proceeds as follows:
\begin{align}
 P_1 \xrightarrow
     [\,\,\,\,\,\,\,\,\,\,\,q, \tau_2\,\,\,\,\,\,\,\,\,\,\,]{T_1\wedge T_2} P_2 \tag{1}\label{Eq:2-sol2} \\
 P_2 \xrightarrow
     [\,\,\,\,\,\,\,\,\,\,\,q, \tau_1\,\,\,\,\,\,\,\,\,\,\,]{T_1} P_1 \tag{2}\label{Eq:2-sol1}
\end{align}
where the rounds satisfy $\tau_1 < \tau_2$. (Hereafter, we assume $\tau_i < \tau_{i+1}$ for any integer $i$.)
$P_1$ first makes a deposit transaction for $T_1 \wedge T_2$. 
Transaction \eqref{Eq:2-sol2} means that $P_2$ can receive $\coins(q)$
only if $P_2$ publishes both $T_1$ and $T_2$ in $\tau_2$.
Namely, it is necessary that both $({\sf Tag}_1, T_1)$ and $({\sf Tag}_2, T_2)$ are valid tag-token pairs to satisfy $\phisr(T_1 \wedge T_2)=1$.
After making the first deposit transaction, $P_2$ makes a deposit transaction for $T_1$. 
Transaction \eqref{Eq:2-sol1} means that $P_1$ can receive $\coins(q)$
only if $P_1$ publishes $T_1$ in $\tau_1$.
In the claim phase, $P_1$ first publishes $T_1$, and then $P_2$ publishes both $T_1$ and $T_2$.

It is important to note that $P_1$ needs to make transaction \eqref{Eq:2-sol2} first.
As a result, $P_2$ cannot claim this transaction without making transaction 
\eqref{Eq:2-sol1} since $P_2$ does not know $T_1$ yet.
Also, the claims are performed in the reverse order of making the transactions,
i.e., $P_1$ first claims.

If $P_2$ aborts after $P_1$ claims, then $P_2$ is penalized with $\coins(q)$ 
and $P_1$ is compensated with that coins.
Thus, $P_2$ needs to publish $T_2$ in order not to lose $\coins(q)$.
Also, both parties never are penalized if they behave honestly.
Therefore, the above protocol satisfies the conditions (A) and (B) in Section \ref{Sect:OV}. 

We show Bentov-Kumaresan's solution for the $n$-party setting in Protocol \ref{Prtcl:BK}.
In the deposit phase, the transactions are created from top to bottom, i.e., \eqref{Eq:BKn-1n} to \eqref{Eq:BKn-21}. 
In the claim phase, the transactions are claimed in the reverse direction, i.e., \eqref{Eq:BKn-21} to \eqref{Eq:BKn-1n}. 
The horizontal lines separate each round. 
Namely, in the deposit (resp. claim) phase, transactions
belonging to the same section are created (resp. claimed) in one round.

Here, we describe an intuitive explanation that Bentov-Kumaresan's fair reconstruction protocol
satisfies the condition (A) and (B).
(See \cite{Bentov14ep} for a formal security proof based on Definition \ref{Def:SCC}.)
It is trivial that no party loses coins if all parties behave honestly.
Thus, we consider the case where there is a party to abort.
 
Let us consider the case where an adversary aborts in the deposit phase.
Since no honest party publishes his/her token, the adversary does not learn the reconstruction
result nor receives any coins from honest parties. This case satisfies the condition (A) and (B).

Let us consider the case where an adversary aborts in the claim phase.
In order to learn the reconstruction result, the adversary must collude all parties that
have not claimed yet to learn tokens that are not published.
Every honest party holds $\coins(q)$ since
he/she has already claimed and has got coins. This case also satisfies the condition (A) and (B).

\subsubsection*{Efficiency}
\begin{figure}[t]
 \begin{center}
 \fbox{\includegraphics[width=12.2cm]{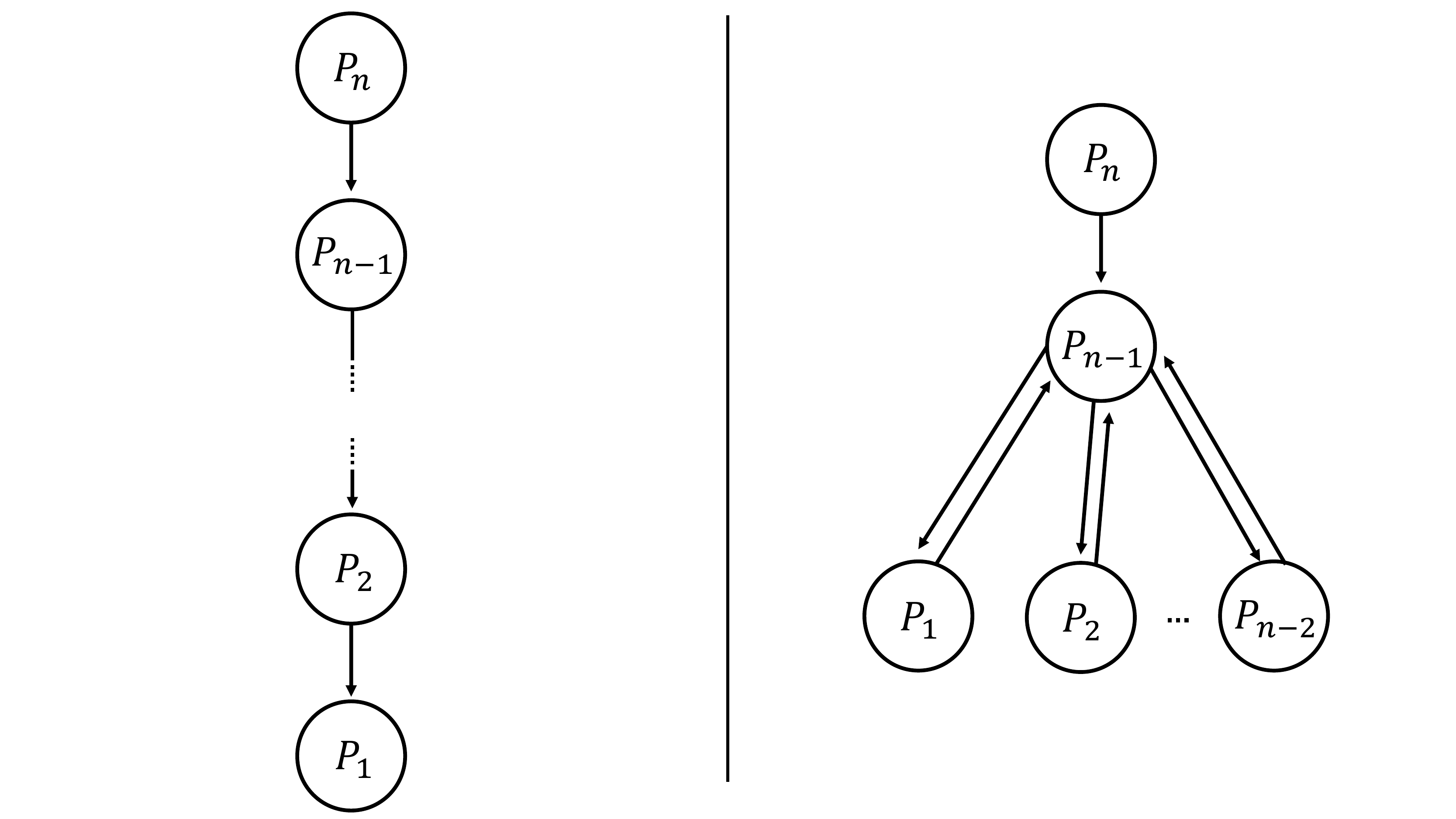}}
  \caption{Flow of Bentov--Kumaresan's fair reconstruction (left) and ours (right)}
  \label{Fig:form}
 \end{center}
\end{figure}

Bentov--Kumaresan's fair reconstruction protocol
requires $n$ rounds for deposit phase and $n$ rounds for claim phase, and thus it requires a total of $2n$ rounds.
(See the left side of Figure \ref{Fig:form} that shows a flow of the claim phase of Bentov--Kumaresan's fair reconstruction protocol.\footnote{For ease of understanding, Figure \ref{Fig:form} omits the transactions (among $P_n$ and other parties) in the last round. (See transactions $(1), (2), \ldots, (n-1)$ in Protocols \ref{Prtcl:BK} and \ref{Prtcl:Ours}.) This also holds for Figure \ref{Fig:redform}.})
Also, it requires $2n-2$ calls of $\Fcr$. 
Recall that the augmented function can be computed in a constant round for any function.
Therefore, for any function, Bentov--Kumaresan's protocol for  
the secure computation with penalties can be {\sf SCC} realized in the 
$(\mathcal{F}_{\mathrm{OT}}, \Fcr)$-hybrid model
with $O(n)$ rounds and $O(n)$ calls of $\Fcr$. 

\begin{spacing}{0.8}
\begin{algorithm}[htbp]
	\floatname{algorithm}{Protocol}
	\caption{Bentov--Kumaresan's Fair Reconstruction Protocol}
	\label{Prtcl:BK}	
	\begin{align}
		P_1 \xrightarrow
		[\,\,\,\,\,\,\,\,\,\,\,\,\,\,\,\,\,\,\,\,\,\,\,\,q, \tau_n\,\,\,\,\,\,\,\,\,\,\,\,\,\,\,\,\,\,\,\,\,\,\,\,]{T_1\wedge \dots \wedge T_n} P_n \tag{1} \label{Eq:BKn-1n}
		\\ 
		P_2 \xrightarrow
		[\,\,\,\,\,\,\,\,\,\,\,\,\,\,\,\,\,\,\,\,\,\,\,\,q, \tau_n\,\,\,\,\,\,\,\,\,\,\,\,\,\,\,\,\,\,\,\,\,\,\,\,]{T_1\wedge \dots \wedge T_n} P_n \tag{2} \label{Eq:BKn-2n}
		\\
		\vdots \,\,\,\,\,\,\,\,\,\,\,\,\,\,\,\,\,\,\,\,\,\,\,\,\,\,\,\,\,\,\nonumber
		\\
		P_{n-1} \xrightarrow
		[\,\,\,\,\,\,\,\,\,\,\,\,\,\,\,\,\,\,\,\,\,\,\,q, \tau_n\,\,\,\,\,\,\,\,\,\,\,\,\,\,\,\,\,\,\,\,\,\,\,\,]{T_1\wedge \dots \wedge T_n} P_n \tag{$n-1$} \label{Eq:BKn-n-1n}
		\\
		\nonumber
		\\
		\noalign{\hrule}\nonumber
		\\
		P_n \xrightarrow
		[\,\,\,\,\,\,\,\,\,\,(n-1)q, \tau_{n-1}\,\,\,\,\,\,\,\,\,\,\,\,]{T_1 \wedge \dots \wedge T_{n-1}} P_{n-1} \tag{$n$} \label{Eq:BKn-nn-1}
		\\
		\nonumber
		\\
		\noalign{\hrule}\nonumber
		\\
		P_{n-1} \xrightarrow
		[\,\,\,\,\,\,\,\,\,\,(n-2)q, \tau_{n-2}\,\,\,\,\,\,\,\,\,\,\,\,]{T_1 \wedge \dots \wedge T_{n-2}} P_{n-2} \tag{$n+1$} \label{Eq:BKn-n-1n-2}
		\\
		\nonumber
		\\
		\noalign{\hrule}\nonumber
		\\
		\vdots \,\,\,\,\,\,\,\,\,\,\,\,\,\,\,\,\,\,\,\,\,\,\,\,\,\,\,\,\,\,\nonumber
		\\
		\nonumber
		\\
		\noalign{\hrule}\nonumber
		\\
		P_{4} \xrightarrow
		[\,\,\,\,\,\,\,\,\,\,\,\,\,\,\,\,\,\,3q, \tau_3\,\,\,\,\,\,\,\,\,\,\,\,\,\,\,\,\,\,\,\,]{T_1 \wedge T_2 \wedge T_3} P_{3} \tag{$2n-4$} \label{Eq:BKn-43}	
		\\
		\nonumber
		\\
		\noalign{\hrule}\nonumber
		\\
		P_{3} \xrightarrow
		[\,\,\,\,\,\,\,\,\,\,\,\,\,\,\,\,\,\,2q, \tau_2\,\,\,\,\,\,\,\,\,\,\,\,\,\,\,\,\,\,\,\,\,]{T_1 \wedge T_2} P_{2} \tag{$2n-3$} \label{Eq:BKn-32}	
		\\
		\nonumber
		\\
		\noalign{\hrule}\nonumber
		\\
		P_{2} \xrightarrow
		[\,\,\,\,\,\,\,\,\,\,\,\,\,\,\,\,\,\, q, \tau_1 \,\,\,\,\,\,\,\,\,\,\,\,\,\,\,\,\,\,\,\,]{T_1} P_{1} \tag{$2n-2$} \label{Eq:BKn-21}	
	\end{align}
\end{algorithm}	
\end{spacing}

\begin{spacing}{0.8}
\begin{algorithm}[htbp]
	\floatname{algorithm}{Protocol}
	\caption{Our Fair Reconstruction Protocol}
	\label{Prtcl:Ours}	
 \begin{align}
  P_1 \xrightarrow
    [\,\,\,\,\,\,\,\,\,\,\,\,\,\,\,\,\,\,\,\,\,\,\,\,\,\,q, \tau_4\,\,\,\,\,\,\,\,\,\,\,\,\,\,\,\,\,\,\,\,\,\,\,\,\,\,]{T_1\wedge \dots \wedge T_n} P_n \tag{1}
     \label{Eq:our-1} \\ 
  P_2 \xrightarrow
    [\,\,\,\,\,\,\,\,\,\,\,\,\,\,\,\,\,\,\,\,\,\,\,\,\,\,q, \tau_4\,\,\,\,\,\,\,\,\,\,\,\,\,\,\,\,\,\,\,\,\,\,\,\,\,\,]{T_1\wedge \dots \wedge T_n} P_n \tag{2}
     \label{Eq:our-2} \\
    \vdots \,\,\,\,\,\,\,\,\,\,\,\,\,\,\,\,\,\,\,\,\,\,\,\,\,\,\,\,\,\,\,\,\nonumber
   \\
  P_{n-1} \xrightarrow
    [\,\,\,\,\,\,\,\,\,\,\,\,\,\,\,\,\,\,\,\,\,\,\,\,\,\,q, \tau_4\,\,\,\,\,\,\,\,\,\,\,\,\,\,\,\,\,\,\,\,\,\,\,\,\,\,]{T_1\wedge \dots \wedge T_n} P_n \tag{$n-1$}
     \label{Eq:our-n-1} \\
  \nonumber
  \\
  \noalign{\hrule}\nonumber
  \\
  P_n \xrightarrow
    [\,\,\,\,\,\,\,\,\,\,\,\,\,\,\,\,\,\,\,\,\,(n-1)q, \tau_3\,\,\,\,\,\,\,\,\,\,\,\,\,\,\,\,\,\,\,\,\,\,\,]{T_1 \wedge \dots \wedge T_{n-1}} P_{n-1} \tag{$n$}
     \label{Eq:our-n} \\
  \nonumber
  \\
  \noalign{\hrule}\nonumber
  \\
  P_{n-1} \xrightarrow
     [\,\,\,\,\,\,\,\,\,\,\,\,\,\,\,\,\,\,\,(n-1)q, \tau_2\,\,\,\,\,\,\,\,\,\,\,\,\,\,\,\,\,\,\,]{T_{n-1} \wedge T_{n-2}} P_{n-2}  \tag{$n+1$}
     \label{Eq:our-n+1} \\
  P_{n-1} \xrightarrow
    [\,\,\,\,\,\,\,\,\,\,\,\,\,\,\,\,\,\,\,(n-1)q, \tau_2\,\,\,\,\,\,\,\,\,\,\,\,\,\,\,\,\,\,\,]{T_{n-1} \wedge T_{n-3}} P_{n-3}  \tag{$n+2$}
     \label{Eq:our-n+2} \\ 
    \vdots \,\,\,\,\,\,\,\,\,\,\,\,\,\,\,\,\,\,\,\,\,\,\,\,\,\,\,\,\,\,\,\,\nonumber
   \\
  P_{n-1} \xrightarrow
    [\,\,\,\,\,\,\,\,\,\,\,\,\,\,\,\,\,\,\,\,\,\,(n-1)q, \tau_2\,\,\,\,\,\,\,\,\,\,\,\,\,\,\,\,\,\,\,\,\,\,]{T_{n-1} \wedge T_1} P_1  \tag{$2n-2$}
     \label{Eq:our-2n-2} \\ 
  \nonumber
  \\
  \noalign{\hrule}\nonumber
  \\
  P_{n-2} \xrightarrow
    [\,\,\,\,\,\,\,\,\,\,\,\,\,\,\,\,(n-2)q, \tau_1\,\,\,\,\,\,\,\,\,\,\,\,\,\,\,\,]{T_{n-1}} P_{n-1}  \tag{$2n-1$}
     \label{Eq:our-2n-1} \\
  P_{n-3} \xrightarrow
    [\,\,\,\,\,\,\,\,\,\,\,\,\,\,\,\,(n-2)q, \tau_1\,\,\,\,\,\,\,\,\,\,\,\,\,\,\,\,]{T_{n-1}} P_{n-1}   \tag{$2n$} 
     \label{Eq:our-2n} \\
    \vdots \,\,\,\,\,\,\,\,\,\,\,\,\,\,\,\,\,\,\,\,\,\,\,\,\,\,\,\,\,\,\,\,\nonumber
   \\
  P_1 \xrightarrow
    [\,\,\,\,\,\,\,\,\,\,\,\,\,\,\,\,\,\, (n-2)q, \tau_1 \,\,\,\,\,\,\,\,\,\,\,\,\,\,\,\,\,\,]{T_{n-1}} P_{n-1}  \tag{$3n-4$}
     \label{Eq:our-3n-4}
 \end{align}
\end{algorithm}	
\end{spacing}


\begin{table}[t]
\centering
\caption{Comparison of fair reconstruction protocols}
   \begin{tabular}{cccc} 
    \toprule
    ~References  \,& $\#$ of Rounds \,& $\#$ of Calling $\Fcr$ \,& Compensations  \\
    \midrule \midrule
    Bentov--Kumaresan \cite{Bentov14} & $2n$ & $2n-2$ & Equivalent  \\
    \midrule
    This work (Sect. \ref{Sect:PP}) & $8$ & $3n-4$ & Non-equivalent  \\
    \bottomrule
   \end{tabular}
    \label{Tbl:fResult}
\end{table}

\section{Proposed Protocol} \label{Sect:PP}
In this section, we introduce a special functionality called secure computation with non-equivalent penalties (Section \ref{Sect:OurPol}). 
Then we design a protocol achieving this functionality in the $(\mathcal{F}_{\mathrm{OT}}, \Fcr)$-hybrid model. 
In particular, we design a new fair reconstruction protocol in the $\Fcr$-hybrid model (Section \ref{Sect:OurFair}), and putting it with a secure computation protocol for an augmented function into the $\mathcal{F}_{\mathrm{OT}}$-hybrid model as in Section \ref{Sect:OV}. 
Notably, our protocol requires $O(1)$ rounds and $O(n)$ broadcasts only (See Table \ref{Tbl:Result}).  

\subsection{Secure computation with non-equivalent penalties} \label{Sect:OurPol}
In secure computation with penalties $\Ff$, all honest parties are compensated with the same amount of money $\coins(q)$. 
A new functionality, secure computation with penalties $\mFf$, is the same as $\Ff$ except that each honest party is compensated with $\coins(q)$ \textit{or more}, i.e., the amount of compensation \textit{may} be different with each party. 
For example, in $\mFf$, we allow the following situation: An honest $P_1$ is compensated with $\coins(q)$ but an honest $P_2$ is compensated with $\coins(2q)$.

See Functionality \ref{Alg:mFf} for a formal definition of $\mFf$. 
The difference with $\Ff$ is that a simulator can decide the amount $q_i$ for each $i\in H'$
and inputs $\coins(\sum_{i \in H'} q_{i})$ while a simulator in $\Ff$ must input $\coins(h'q)$
for $h' := |H'|$. 
We require that $q_i \geq q$ for all $i \in H'$, where $q$ is the minimum amount of compensation.

We note that compensation happens only when a malicious party has stolen the output value. 
That is, $\Ff$ and $\mFf$ are the same if all parties behave honestly. 
By choosing $q$ appropriately, it is possible to prevent malicious behavior, and then we obtain a protocol with fairness. 
In this sense, a new functionality $\mFf$ brings almost the same effect on $\Ff$.

\subsection{Fair reconstruction for secure computation with non-equivalent penalties} \label{Sect:OurProcedure}
Following Bentov-Kumaresan's protocol,
we construct a fair reconstruction protocol to realize 
secure computation with non-equivalent penalties.
In order to realize secure computation with non-equivalent penalties,
a fair reconstruction protocol needs to satisfy the following
conditions:
\begin{itemize}
	\item[(A)]  No honest party pays any penalty.
	\item[(B*)] If an adversary learns the reconstruction result, 
	but an honest party cannot,
	then the honest party is compensated with coins.
	Furthermore, the compensation is more than a predetermined amount.
\end{itemize}
Note that the difference between condition (B*) and condition (B) in Section \ref{Sect:OV} is the amount of compensations only. 
Namely, our fair reconstruction protocol does not guarantee that each honest party is compensated with the same amount of coins.

\subsection{Our fair reconstruction protocol}\label{Sect:OurFair}

Our fair reconstruction protocol 
proceeds as follows 
(see also Protocol \ref{Prtcl:Ours} that shows the protocol expressed by using the arrow notation.)\footnote{Protocol \ref{Prtcl:Ours} is described based on the same rules as protocol \ref{Prtcl:BK}. Namely, the transactions are created from top to bottom and claimed in the reverse direction. The horizontal lines separate each round.}:

\medskip
\noindent
\textbf{Deposit phase:}
\begin{enumerate}
		\item For $i \in \{1, \dots, n-1\}$, $P_i$ makes a transaction to send $P_n$ $\coins(q)$
		with a predicate $\phi_{i,n}$ and a round number $\tau_4$,
		where $\phi_{i,n}(x) = 1$ only if $x = T_1 \wedge \dots \wedge T_n$.
		\item $P_n$ makes a transaction to send $P_{n-1}$ $\coins((n-1)q)$
		with a predicate $\phi_{n,n-1}$ and a round number $\tau_3$,
		where $\phi_{n,n-1}(x) = 1$ only if $x = T_1 \wedge \dots \wedge T_{n-1}$.
		\item For $i \in \{1,\dots,n-2\}$,
		$P_{n-1}$ makes a transaction to send $P_i$ $\coins((n-1)q)$
		with a predicate $\phi_{n-1,i}$ and a round number $\tau_2$,
		where $\phi_{n-1,i}(x) = 1$ only if $x = T_{n-1} \wedge T_i$.
		\item For $i \in \{1,\dots,n-2\}$, $P_i$ makes a transaction to send $P_{n-1}$
		$\coins((n-2)q)$
		with a predicate $\phi_{i,n-1}$ and a round number $\tau_1$,
		where $\phi_{i,n-1}(x) = 1$ only if $x = T_{n-1}$.
	\end{enumerate}
\medskip
\noindent
\textbf{Claim phase:}
	\begin{enumerate}	
	\setcounter{enumi}{4}
		\item $P_{n-1}$ claims by publishing $T_{n-1}$ in round $\tau_1$ and receives $\coins((n-2)q)$ 
		from each of $P_1, \dots, P_{n-2}$.
		\item For $i \in \{1,\dots,n-2\}$,
		$P_i$ claims by publishing $T_{n-1} \wedge T_i$ in round $\tau_2$ and receives 
		$\coins((n-1)q)$ from $P_{n-1}$.
		\item $P_{n-1}$ claims by publishing $T_1 \wedge \dots \wedge T_{n-1}$
		in round $\tau_3$ and receives $\coins((n-1)q)$ from $P_n$.
		\item $P_n$ claims by publishing $T_1 \wedge \dots \wedge T_n$ in round $\tau_4$
		and receives $\coins(q)$ from each of $P_1, \dots, P_{n-1}$.
\end{enumerate}

Our fair reconstruction protocol requires eight rounds and $3n-4$ calls of $\Fcr$.
Since $\mathcal{F}_{\mathrm{OT}}$ is sufficient to compute any standard
functionality in constant rounds, we can derive the following theorem.
 
\medskip
\noindent
\begin{thm}\label{Thm:Scc}
Assuming the existing of one-way functions, for every $n$-party functionality $f$
there exists a protocol that {\sf SCC} realizes $\mFf$
in the $(\mathcal{F}_{\mathrm{OT}}, \Fcr)$-hybrid model.
The protocol requires $O(1)$ rounds and $O(n)$ calls of $\Fcr$.
\end{thm}

\subsection{Idea behind our protocol}
See the right side of Figure \ref{Fig:form} that shows a flow of the claim phase
of our fair reconstruction protocol.
In Bentov--Kumaresan's protocol, 
parties publishes his/her token in serial order, i.e., 
each token is published in each round.
(Token $T_i$ is published in round $\tau_i$.)
Thus, their protocol requires $O(n)$ rounds.

On the other hand, 
our protocol enables to publish multiple tokens in one round
to improve the round complexity.
See step 6) in Section \ref{Sect:OurFair}, the parties $P_1, \dots, P_{n-2}$ 
publish their token in one round.

In the claim phase, our protocol proceeds as follows:
We call $P_1, \dots, P_{n-2}$ \textit{middle parties}  and
$P_{n-1}$ \textit{aggregator}.
In round $\tau_1$, 
the aggregator $P_{n-1}$ collects coins from all middle parties by publishing token $T_{n-1}$.
After that, the middle parties publishes their tokens $T_1, \dots, T_{n-2}$
and receive coins, which are more than they sent in round $\tau_1$, 
from the aggregator $P_{n-1}$ in round $\tau_2$.
In round $\tau_3$, the aggregator $P_{n-1}$ receives coins from $P_n$ 
by publishing his/her token and all of middle parties' tokens.
In the last round $\tau_4$, $P_n$ publishes the last token $T_n$ and receives coins from
every other party.
As a result, all parties learn the reconstruction result and every party's wallet
are balanced, i.e., it has neither loss nor gain.

We discuss the amount of coins sent in each transaction to satisfy the conditions
(A) and (B*) below.

\subsubsection*{The amount of coins}
\begin{figure}[p]
 \begin{center}
  \fbox{\includegraphics[width=12.2cm]{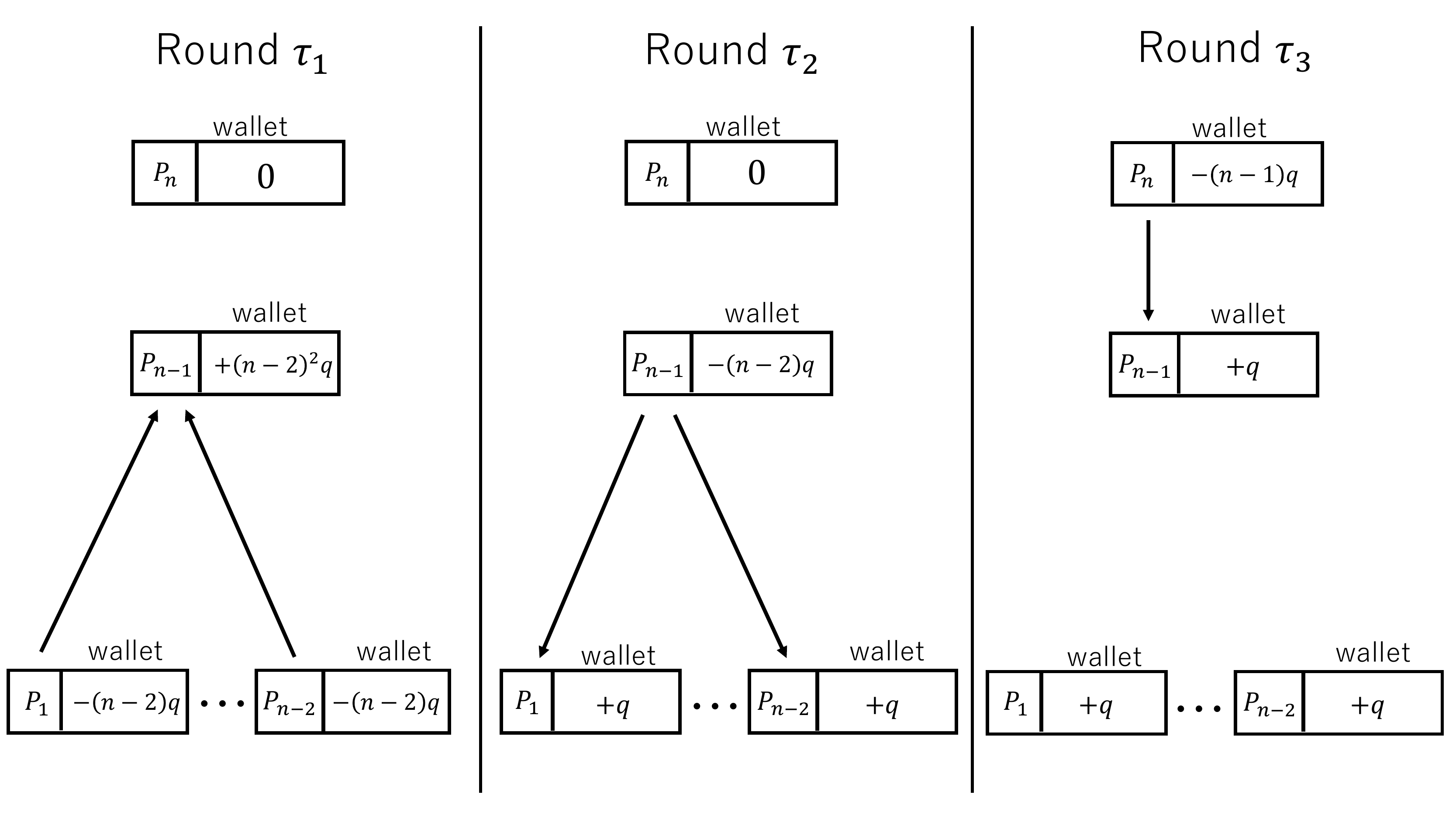}}
  \caption{Coins flow in round $\tau_1$ to $\tau_3$ in the case where all parties behave honestly}
  \label{Fig:honest}
 \end{center}
\end{figure}

In our protocol, $P_n$ receives $\coins(q)$ 
from every other party in the last round $\tau_4$.
In order to satisfy the condition (A),
every wallet of $P_1, \dots, P_{n-1}$ must hold $\coins(q)$ at the end of
round $\tau_3$.
We show that our protocol satisfies this condition in Figure \ref{Fig:honest}.

When we decide the amount of coins in rounds $\tau_1$ and $\tau_2$,
we should note that the aggregator $P_{n-1}$ cannot claim in round $\tau_3$ if
at least one of the middle parties abort in round $\tau_2$.
Since the aggregator sends more coins in round $\tau_2$ than he/she received
in round $\tau_1$, his/her wallet holds negative amount of coins at the end of round $\tau_2$.
In order to satisfy the conditions (A) and (B*), 
it is necessary to satisfy that the aggregator's wallet holds positive amount of
coins at the end of round $\tau_2$ if at least one of the middle parties abort in round $\tau_2$. 
The amounts of coins sent in rounds $\tau_1$ and $\tau_2$ are derived as follows.

Suppose that $P_{n-1}$ gets $\coins(xq)$ from each of $P_1, \dots, P_{n-2}$ in round $\tau_1$, and each of $P_1, \dots, P_{n-2}$ get $\coins((x+1)q)$ from $P_{n-1}$ in round $\tau_2$. 
In round $\tau_2$, $P_{n-1}$'s wallet should have positive amount of coins unless all of $P_1, \dots, P_{n-2}$ claims. 
Thus, we can derive $x$ from the following equation: 
$(n-2)x > (n-3)(x+1)$.
The least solution of the equation is $x = n-2$. 
Therefore, each middle party sends $\coins((n-2)q)$ to the aggregator 
in round $\tau_1$ and the aggregator sends $\coins((n-1)q)$ to each middle party in round $\tau_2$.

\subsubsection*{Security intuition}
\begin{figure}[p]
	\begin{center}
		\fbox{\includegraphics[width=12.2cm]{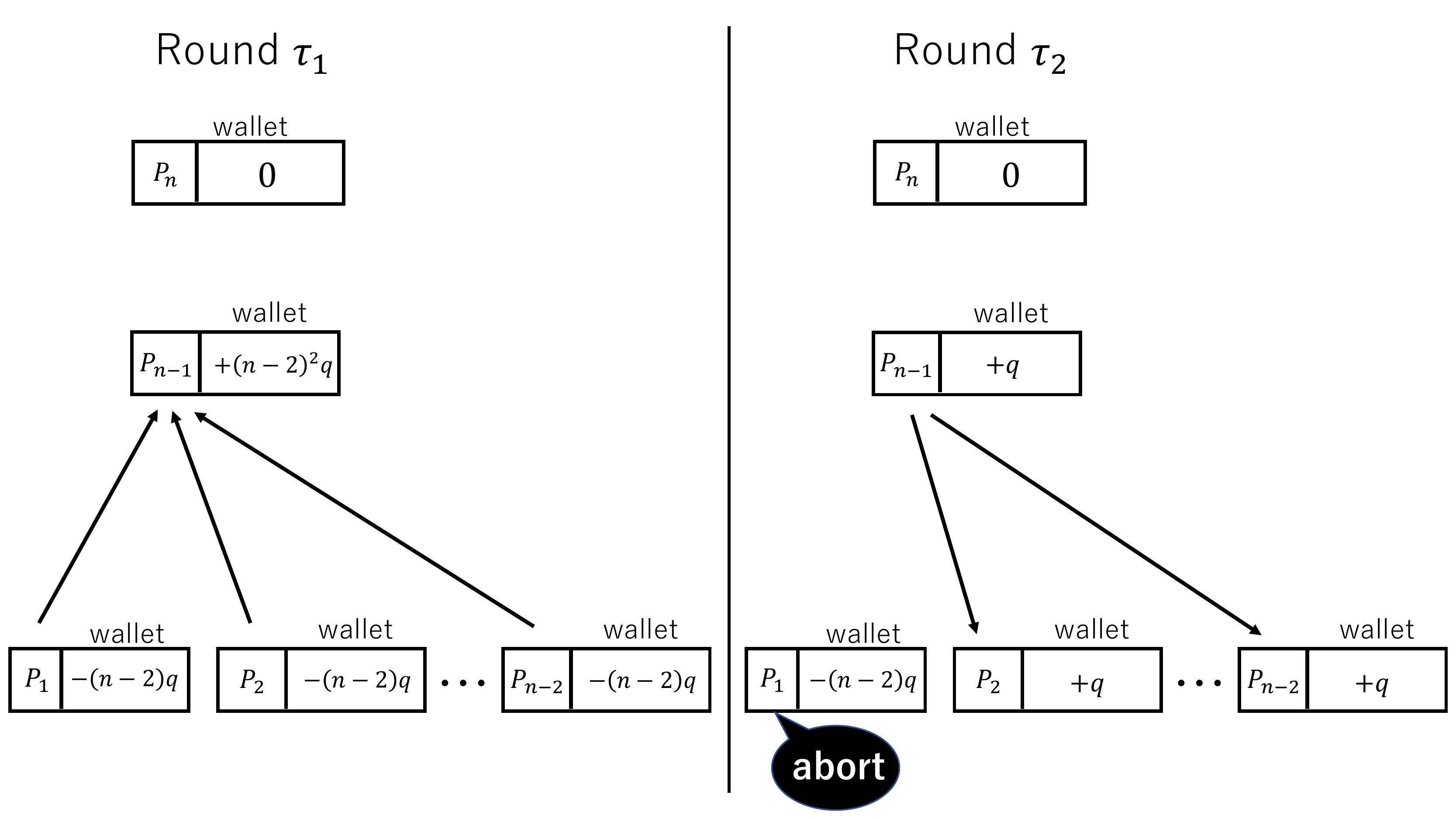}}
		\caption{Coins flow in round $\tau_1$ to $\tau_2$ in the case where $P_1$ aborts}
		\label{Fig:abort}
	\end{center}
\end{figure}

Let us consider the case where one of the middle parties aborts in round $\tau_2$.
(See Figure \ref{Fig:abort}.)
Suppose that $P_1$ aborts in round $\tau_2$, i.e., he/she does not publish
$T_1$ and does not receive $\coins((n-1)q)$ from $P_{n-1}$.
Note that $P_1$ must collude with $P_n$ to learn the reconstruction result.
Thus, the condition (B*) is satisfied since
every wallet of $P_2, \dots, P_{n-1}$ holds $\coins(q)$ as the compensation
at the end of the protocol.
Furthermore, since no honest party does not pay a penalty, the condition (A) is satisfied.
We can confirm that our protocol satisfies the conditions (A) and (B*)
by the same way in the other cases.

\subsubsection*{Remark 1}
Compensations to honest parties may not be the same amount of coins. 
See $P_{n-1}$ who receives $\coins((n-2)q)$ from each of $P_1, \dots, P_{n-2}$
 in round $\tau_1$.
The amount of $P_{n-1}$'s compensation depends on the number of aborting parties in them.
On the other hand, compensations for other parties are $\coins(q)$.
Namely, $P_{n-1}$ is the only party who can be compensated with more than $\coins(q)$.

\subsubsection*{Remark 2}
At first glance, it seems that rounds $\tau_3$ and $\tau_4$ need not be
separated since $P_n$ can already claim in $\tau_3$.
However, if these rounds are combined into one (i.e., $\tau_4 = \tau_3$),
the modified protocol violates condition (A). 
Suppose all but $P_n$ are malicious. 
First, in the deposit phase, the adversary makes the $n-1$ transactions to $P_n$ honestly. 
However, after $P_n$ makes the deposit transaction to $P_{n-1}$, the adversary waits 
for time to pass without making the subsequent transactions.
Just before the end of $\tau_3$, the adversary claims the transaction made by $P_n$
and obtains $\coins((n-1)q)$.
$P_{n-1}$ can get that coins back by claiming $n-1$ transactions made by the 
adversary, however $P_n$ may not claim due to the lack of time remaining.
As a result, $P_n$ may lose the coins, which violates the condition (A).

\begin{algorithm}[htbp]
\floatname{algorithm}{Functionality}
 \caption{Secure computation with non-equivalent penalties $\mFf$}
 \label{Alg:mFf}
 \begin{description}
  \item[\textbf{Setup}]  \,\\
  The session identifier is $sid$. 
  Running with parties $P_1, \dots, P_n$, 
  and an ideal adversary $\mathcal{S}$ that corrupts parties $\{P_i\}_{i\in C}$. 
  Let $d$ be a parameter representing the safety deposit. 
  Let $q$ denote the minimum penalty amount.
 \item[\textbf{Input phase}]  \,\\
 Wait to receive the following messages.
   \renewcommand{\labelitemi}{-}
 \begin{itemize}
  \item  $(\mathrm{input}, sid, ssid, i, x_i, \coins(d))$ from $P_i$ for all $i \in H$
  \item  $(\mathrm{input}, sid, ssid, \{x_i\}_{i \in C}, H', \coins(\sum_{i \in H'} q_{i}))$ 
         from $\mathcal{S}$, where $H' \subseteq H$ and $q_{i}\,(\geq q)$ is
	 the penalty amount for each $i \in H'$.
 \end{itemize}
 \item[\textbf{Output phase}] \,\\
 Perform the following process.
 \begin{enumerate}
  \item Send $(\mathrm{return}, sid, ssid, \coins(d))$ to each $P_r$ for $r \in H$.
  \item Compute $(y_1, \dots, y_n) \leftarrow f(x_1, \dots, x_n)$.
  \renewcommand{\labelitemi}{-}
  \begin{itemize}
   \item if $h' =0$, then send message $(\mathrm{output}, sid, ssid, z_r)$ to $P_r$ for 
         $r \in H$, and terminate.
   \item If $0< h' <h$, then send $(\mathrm{extra}, sid, ssid, \coins(q_i))$ to $P_i$ 
	 for each $i \in H'$, and terminate, where $h := |H|$.
   \item If $h' = h$, then send message $(\mathrm{output}, sid, ssid, \{y_i\}_{i \in C})$
	 to $\mathcal{S}$.
  \end{itemize}
  \item If $\mathcal{S}$ returns $(\mathrm{continue}, sid, ssid, H'')$, 
	where $H'' \subseteq H$,
	then perform the following process:
	\begin{enumerate}
	 \setlength{\leftskip}{0.5cm}
	 \item send $(\mathrm{output}, sid, ssid, y_i)$ to $P_i$ 
               for all $i \in H$,
	 \item send $(\mathrm{payback}, sid, ssid, \coins(\sum_{i \in H'} q_{i} - \sum_{j \in H''} q_j)$ 
	       to $\mathcal{S}$ where $h'' = |H''|$,
	 \item send $(\mathrm{extrapay}, sid, ssid, \coins(q_i))$ to $P_i$ for each 
	       $i \in H''$.
	\end{enumerate}
  \item Else if $\mathcal{S}$ returns $(\mathrm{abort}, sid, ssid)$, send
	$(\mathrm{penalty}, sid, ssid, \coins(q_i))$ to $P_i$ for each $i \in H$.
 \end{enumerate}
 \end{description}
\end{algorithm}

\begin{figure}[ht]
	\begin{center}
		\fbox{\includegraphics[width=12.2cm]{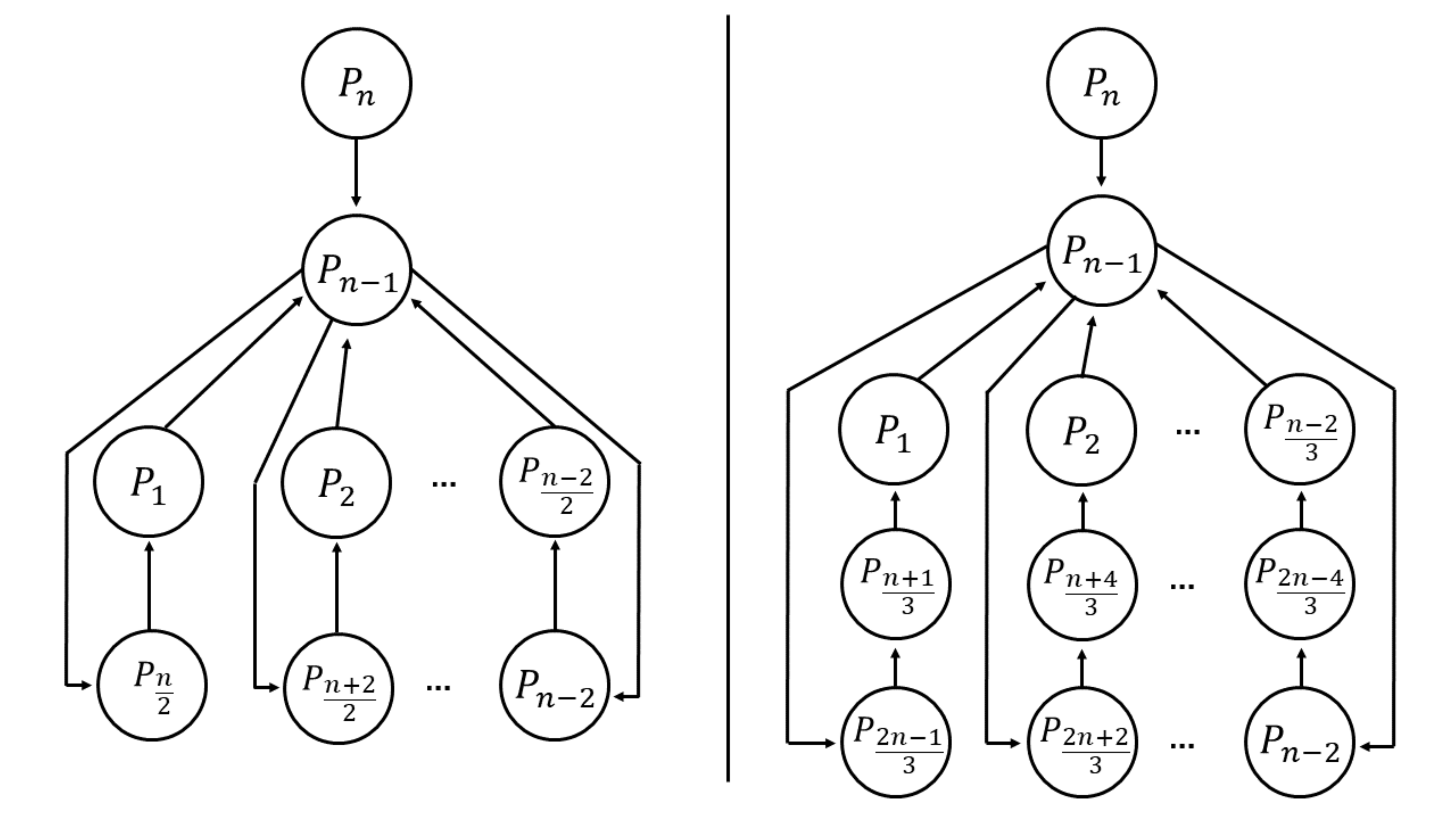}}
		\caption{Our fair reconstruction in the case where $l=1$ (left) and $l=2$ (right)}
		\label{Fig:redform}
	\end{center}
\end{figure}

\section{Reducing Deposit Amount} \label{Sect:Reduce}
This section focuses on the amount of coins required to run our protocol.
Our protocol requires larger deposits than Bentov-Kumaresan’s protocol.
This section shows a technique to reduce the deposit amount of our protocol.

\subsection{Deposit amount}
We first discuss the largest amount of coins for participation. 
In Bentov-Kumaresan's protocol, 
$P_n$ needs to deposit $\coins((n-1)q)$ to $P_{n-1}$,
which is the largest amount in all parties. 
Thus, he/she must hold $\coins((n-1)q)$ at the beginning
of the protocol to participate in the protocol.

On the other hand, in our protocol, 
$P_{n-1}$ needs to make deposit transactions $(n-1)$ and $(n+1)$ to $(2n-2)$.
Thus, he/she must hold the total amount $\coins(((n-1)(n-2)+1)q)$
at the beginning of the protocol.
Although our protocol is efficient in terms of 
the round complexity, the deposit amount increases from $O(nq)$ to $O(n^2q)$.

\subsection{Reducing the deposit amount} 
We present a technique of reducing the deposit amount 
in our protocol.
Our technique can reduce $P_{n-1}$'s deposit to $\coins(((n-1)(n-2)/(l+1) +1)q)$ instead of increasing the number of rounds by $2l$. (Each of the deposit phase and the claim phase is increased by $l$ rounds.)

The reason for $P_{n-1}$'s large deposit is that
he/she has to make deposit transactions for all of the 
middle parties, i.e., $P_1, \dots, P_{n-2}$.
Our main idea is to reduce the number of deposit transactions to the middle parties
from $P_{n-1}$.
Figure \ref{Fig:redform} shows flows of our protocol applied the above idea in the case of $l=1$ and $l=2$. 
(See also Protocol \ref{Prtcl:FRPred}, which shows the specific process in the case where $l=1$.)
In order to reduce $P_{n-1}$'s deposit,
the middle parties make deposit transactions within them.
For instance, when $l=1$, the number of deposit transactions that
$P_{n-1}$ makes for the middle parties is halved instead of increasing the number of rounds by two.
As a result, $P_{n-1}$'s deposit is reduced to $\coins((n-1)(n-2)/2 +1)q)$.
Similarly, when $l=2$, his/her deposit is reduced to $1/3$ instead of increasing the number of rounds by four.
We remark that our technique can be applied even if $n-2$ is not divided by $l+1$ by having one of the middle parties be responsible for multiple middle parties. 



\begin{spacing}{0.1}
\begin{algorithm}[htbp]
\floatname{algorithm}{Protocol}
\caption{Reducing the deposit amount (case of $l=1$)}
\label{Prtcl:FRPred}	
\floatname{algorithm}{Protocol}	
 \begin{align}
 		P_1 \xrightarrow
		[\,\,\,\,\,\,\,\,\,\,\,\,\,\,\,\,\,\,\,\,\,\,\,\,\,\,\,\,\,\,q, \tau_5\,\,\,\,\,\,\,\,\,\,\,\,\,\,\,\,\,\,\,\,\,\,\,\,\,\,\,\,\,\,]{T_1\wedge \dots \wedge T_n} P_n \tag{1} \label{Eq:redeven-n-1n}
		\\ 
		\vdots \,\,\,\,\,\,\,\,\,\,\,\,\,\,\,\,\,\,\,\,\,\,\,\,\,\,\,\,\,\,\nonumber
		\\
		P_{n-1} \xrightarrow
		[\,\,\,\,\,\,\,\,\,\,\,\,\,\,\,\,\,\,\,\,\,\,\,\,\,\,\,\,\,q, \tau_5\,\,\,\,\,\,\,\,\,\,\,\,\,\,\,\,\,\,\,\,\,\,\,\,\,\,\,\,\,\,]{T_1\wedge \dots \wedge T_n} P_n \tag{$n-1$}
		\label{Eq:Eq:redeven-n-1n}
		\\
		\nonumber
		\\
		\noalign{\hrule} \nonumber
		\\
		P_n \xrightarrow
		[\,\,\,\,\,\,\,\,\,\,\,\,\,\,\,\,(n-1)q, \tau_{4}\,\,\,\,\,\,\,\,\,\,\,\,\,\,\,\,\,\,]{T_1 \wedge \dots \wedge T_{n-1}} P_{n-1} \tag{$n$} \label{Eq:Eq:redeven-nn-1}
		\\
		\nonumber
		\\
		\noalign{\hrule}\nonumber
		\\
  P_{n-1} \xrightarrow
     [\,\,\,\,\,\,\,\,\,\,\,\,\,\,\,\,\,\,\,\,\,\,\,\,\,(n-1)q, \tau_3\,\,\,\,\,\,\,\,\,\,\,\,\,\,\,\,\,\,\,\,\,\,\,\,\,]{T_{n-1} \wedge T_{n-2} \wedge T_{n/2}} P_{n/2-1}  \tag{$n+1$}
     \label{Eq:redeven-n-1n-1} \\
    \vdots \,\,\,\,\,\,\,\,\,\,\,\,\,\,\,\,\,\,\,\,\,\,\,\,\,\,\,\,\,\,\,\,\nonumber
   \\
  P_{n-1} \xrightarrow
    [\,\,\,\,\,\,\,\,\,\,\,\,\,\,\,\,\,\,\,\,\,\,\,\,\,\,\,\,(n-1)q, \tau_3\,\,\,\,\,\,\,\,\,\,\,\,\,\,\,\,\,\,\,\,\,\,\,\,\,\,\,\,]{T_{n-1} \wedge T_{n/2 +1} \wedge T_1} P_1  \tag{$3n/2-1$}
     \label{Eq:redeven-n-12} \\ 
  \nonumber
  \\
  \noalign{\hrule}\nonumber
  \\
  P_{n/2-1} \xrightarrow
    [\,\,\,\,\,\,\,\,\,\,\,\,\,\,\,\,\,\,\,\,\,\,(n-2)q, \tau_2\,\,\,\,\,\,\,\,\,\,\,\,\,\,\,\,\,\,\,\,\,\,]{T_{n-1}\wedge T_{n-2}} P_{n-2}  \tag{$3n/2$}
     \label{Eq:redeven-n/2-1-n-1} \\
    \vdots \,\,\,\,\,\,\,\,\,\,\,\,\,\,\,\,\,\,\,\,\,\,\,\,\,\,\,\,\,\,\,\,\nonumber
   \\
  P_1 \xrightarrow
    [\,\,\,\,\,\,\,\,\,\,\,\,\,\,\,\,\,\,\,\,\,\,\,\, (n-2)q, \tau_2 \,\,\,\,\,\,\,\,\,\,\,\,\,\,\,\,\,\,\,\,\,\,\,\,]{T_{n-1}\wedge T_{n/2 +1}} P_{n/2 +1}  \tag{$2n-2$} 
     \label{Eq:redeven-1-n-1}  \\
       \nonumber
  \\
  \noalign{\hrule}\nonumber
  \\
  P_{n-2} \xrightarrow
    [\,\,\,\,\,\,\,\,\,\,\,\,\,\,\,\,\,\,\,\,\,\,(n-3)q, \tau_1\,\,\,\,\,\,\,\,\,\,\,\,\,\,\,\,\,\,\,\,\,\,]{T_{n-1}} P_{n-1}  \tag{$2n-1$}
     \label{Eq:redeven-n-2n-1} \\
    \vdots \,\,\,\,\,\,\,\,\,\,\,\,\,\,\,\,\,\,\,\,\,\,\,\,\,\,\,\,\,\,\,\,\nonumber
   \\
  P_{n/2+1} \xrightarrow
    [\,\,\,\,\,\,\,\,\,\,\,\,\,\,\,\,\,\,\,\,\,\,\,\, (n-3)q, \tau_1 \,\,\,\,\,\,\,\,\,\,\,\,\,\,\,\,\,\,\,\,\,\,\,\,]{T_{n-1}} P_{n-1}  \tag{$5n/2-4$} 
     \label{Eq:redeven-n/21-n-1}
 \end{align}
\end{algorithm}	
\end{spacing}

\begin{algorithm}[htbp]
	\floatname{algorithm}{Functionality}
	\caption{Claim-refund-or-give functionality $\Fcrg$}
	\label{Alg:Fcrg}
	\begin{description}
		\item[\textbf{Setup}:] The session identifier is $sid$. 
		Running with parties $P_1, \dots, P_n$ and an ideal adversary $\mathcal{S}$.
		\vspace{-2mm}
		\item[\textbf{Deposit phase}:]
		Receiving $(\mathrm{deposit}, sid, ssid, s, r, \phisr, \phirf, \tau_{\mathrm{claim}}, \tau_{\mathrm{refund}}, \coins(c))$ from $P_s$, and $\phirf'$ from $P_r$,
		perform the following process.
		\begin{enumerate}
		    \vspace{-2mm}
		    \item Check $\phirf = \phirf'$ holds.
			\item Record the message $(\mathrm{deposit}, sid, ssid, s, r, \phisr, \phirf, \tau_{\mathrm{claim}}, \tau_{\mathrm{refund}}, c)$
			\item Send all parties $(\mathrm{deposit}, sid, ssid, s, r, \phisr, \phirf, \tau_{\mathrm{claim}}, \tau_{\mathrm{refund}}, c)$
			\renewcommand{\labelitemi}{-}
			\begin{itemize}
				\item Ignore any future messages with the same $ssid$ from $P_s$ to $P_r$.
			\end{itemize}
		\end{enumerate}
		\vspace{-4mm}
		\item[\textbf{Claim phase}:]
		Receiving $(\mathrm{claim}, sid, ssid, s, r, \phisr, \phirf, \tau_{\mathrm{claim}}, \tau_{\mathrm{refund}}, c, w_r)$ 
		from $P_r$ in round $\tau_{\mathrm{claim}}$, perform the following process.
		\begin{enumerate}
		    \vspace{-2mm}
			\item Check that both $(\mathrm{deposit}, sid, ssid, s, r, \phisr, \phirf, \tau_{\mathrm{claim}}, \tau_{\mathrm{refund}}, c)$ was recorded, and $\phisr(w_r) = 1$ hold. 
			\item send $(\mathrm{claim}, sid, ssid, s, r, \phisr, \phirf, \tau_{\mathrm{claim}}, \tau_{\mathrm{refund}}, c, w_r)$ to all parties, and $(\mathrm{claim}, sid, ssid, s, r, \phisr, \phirf, \tau_{\mathrm{claim}}, \tau_{\mathrm{refund}}, \coins(c))$ to $P_r$,
			\item delete the record $(\mathrm{deposit}, sid, ssid, s, r, \phisr, \phirf, \tau_{\mathrm{claim}}, \tau_{\mathrm{refund}}, c)$.
		\end{enumerate}
		\vspace{-4mm}
		\item[\textbf{Refund phase}:] 
		In $\tau_{\mathrm{claim}} + 1$, if the record $(\mathrm{deposit}, sid, ssid, s, r, \phisr, \phirf, \tau_{\mathrm{claim}}, \tau_{\mathrm{refund}}, c)$ was not deleted and the message $(\mathrm{refund}, sid, ssid, s, r, \phisr, \phirf, \tau_{\mathrm{claim}}, \tau_{\mathrm{refund}}, c, w_r)$ is received from $P_r$, then perform the following process:
		\vspace{-2mm}
		\begin{enumerate}
			\item Check that both $(\mathrm{deposit}, sid, ssid, s, r, \phisr, \phirf, \tau_{\mathrm{claim}}, \tau_{\mathrm{refund}}, c)$ was recorded,
			and $\phirf(w_r) = 1$ hold.
			\item Send $(\mathrm{claim}, sid, ssid, s, r, \phisr, \phirf, 		   \tau_{\mathrm{claim}}, \tau_{\mathrm{refund}}, c, w_s)$ to all parties, and
			$(\mathrm{refund}, sid, ssid, s, r, \phisr, \phirf, \tau_{\mathrm{claim}}, \tau_{\mathrm{refund}}, \coins(c))$ to $P_s$.
			\item Delete the record $(\mathrm{deposit}, sid, ssid, s, r, \phisr, \phirf, \tau_{\mathrm{claim}}, \tau_{\mathrm{refund}}, c)$.
		\end{enumerate}
		\vspace{-4mm}
		\item[\textbf{Give phase}:]
		In $\tau_{\mathrm{refund}} + 1$, if the record $(\mathrm{deposit}, sid, ssid, s, r, \phisr, \phirf, \tau_{\mathrm{claim}}, \tau_{\mathrm{refund}}, c)$ was not deleted, 
		then perform the following process:
		\begin{enumerate}
			\vspace{-2mm}
			\item Send $(\mathrm{give}, sid, ssid, s, r, \phisr, \phirf,\tau_{\mathrm{claim}}, \tau_{\mathrm{refund}}, \coins(c))$ to $P_r$.
			\item Delete the record $(\mathrm{deposit}, sid, ssid, s, r, \phisr, \phirf, \tau_{\mathrm{claim}}, \tau_{\mathrm{refund}}, c)$.
		\end{enumerate}
	\end{description}
\end{algorithm}

\section{Achieving Equivalent Compensation} \label{Sect:Equiv}
In this section, we describe how to achieve secure computation with equivalent penalties with $O(1)$ rounds and $O(n)$ broadcasts.

We first recall the reason for the non-equivalence. 
$P_{n-1}$'s compensation amount depends on the number of parties who aborts in step 6). (See Section \ref{Sect:OurProcedure}.)
Let $x \in [n-2]$ be the number of parties who abort in step 6).
Then, $P_{n-1}$'s compensation amount is $\coins(((x-1)n+2-x)q) = \coins(((n-2)^2 - (n-2-x)(n-1))q)$.
(Note that $n-2-x$ is the number of parties who claim in step 6).)
Thus, $P_{n-1}$ is compensated with more than $\coins(q)$ if two or more parties abort in step 6), although compensation amounts for the other parties are always $\coins(q)$.
In order to remove the non-equivalence,
it is necessary that $P_{n-1}$ is compensated with the same amount of coins as the other honest parties even if two or more parties abort in step 6).

\subsection{A solution based on the honesty of the parties}
A simple solution is that $P_{n-1}$ distributes coins to the other honest parties so that the compensation amounts are equivalent.
More precisely, when $x\, (\geq 2)$ parties abort in step 6),
$P_{n-1}$ transfers coins\footnote{Since the sum of the compensations is $\coins((n-2)xq) = \coins(((x-1)n+2-x)q+(n-2+x)q)$, $P_{n-1}$ sends $\coins((n-2)xq/(n-1+x))$ to each of parties who claim in step 6). }
$n-2-x$ parties who claim in step 6)
to make the compensations equivalent for $P_{n-1}$ and them,
after terminating the protocol.

However, this solution has no enforcement for $P_{n-1}$ to distribute coins since $P_{n-1}$ is imposed no penalty if he/she refuses to distribute coins.
It is undesirable since it relies too much on $P_{n-1}$'s honest behaviour.
In the below, we present a technique of achieving the equivalent
compensation in a more enforceable way.

\subsection{Claim-refund-or-give functionality}
The main idea of satisfying the equivalent compensation
is to build a mechanism whereby $P_{n-1}$ cannot refund for more than one transaction in step 6).
For instance, when $P_1$ and $P_2$ abort in step 6), the mechanism forces $P_{n-1}$ to only get $\coins((n-1)q)$ back. 
Namely, if $P_{n-1}$ gets back $\coins((n-1)q)$ from the transaction made for $P_1$,
then $P_{n-1}$ is forced to \textit{give} $\coins((n-1)q)$ to $P_2$.
As a result, $P_{n-1}$ is compensated with $\coins(q)$ as the other honest parties.

To put the above idea, we introduce a new functionality called \textit{claim-refund-or-give functionality} $\Fcrg$, which is an extension of the claim-or-refund functionality.
$\Fcrg$ is the same as $\Fcr$ except that there is a condition for a sender to get the coins back in the refund phase.
When a receiver does not reveal the witness in the claim phase,
a sender can refund only if the sender reveals his/her witness $w_s$.
If the sender does not reveal the witness $w_s$ in the refund phase, 
the sender is forced to give the receiver the coins.

$\Fcrg$ consists of four phases: deposit, claim, refund, and give. (See Functionality \ref{Alg:Fcrg} for a formal description of $\Fcrg$.)
In the deposit phase, a sender $P_s$ sends conditional coins with two predicates
$\phisr, \phirf$ and two round numbers $\tau_{\mathrm{claim}}, \tau_{\mathrm{refund}}$
where $\tau_{\mathrm{claim}} < \tau_{\mathrm{refund}}$.
(We note that, in our protocol, $\phisr$ is determined by the sender, but
$\phirf$ is collaboratively designed with the sender and the receiver.)
In the claim phase, the receiver $P_r$ gets the coins only if he/she submits a witness $w_r$ such as $\phisr(w_r)=1$ within $\tau_{\mathrm{claim}}$.
In the refund phase, the sender $P_s$ gets back the coins only if
$P_r$ does not claim in $\tau_{\mathrm{claim}}$ and $P_s$ 
reveals a witness $w_s$ such as $\phirf(w_s)=1$ within $\tau_{\mathrm{refund}}$.
In the give phase, if $P_s$ does not claim and $P_r$ does not refund, 
$P_r$ receives the coins.
Note that $\Fcrg$ is the same as $\Fcr$ if $\phirf$ is a tautology since $P_r$ can refund unconditionally in the refund phase.

\subsection{How to resolve the non-equivalence}
Using $\Fcrg$,  we construct a mechanism to prevent $P_{n-1}$ refunds two or more transactions in step 6).
Our solution is as follows:
In the first phase, i.e., secure computation for the augmented function $\hat f$,
the parties generate the following values in addition to the secret shares of $\hat f$.
\begin{itemize}
	\item $(S_1,\dots,S_{n-2})$: the token $S_i$ is a pair $(s_i,d_i)$ of a secret share $s_i$ of the $P_{n-1}$'s witness $w$, and the de-commitment $d_i$ of the secret share.
	      Here, the witness $w$ is chosen
	      from $\{0,1\}^k$ uniformly at random.
	      The underlying secret sharing scheme is {\sf pubNMSS} where the threshold value is \textit{two}. Only $P_{n-1}$ receives all of these values.
	\item $(c_1,\dots, c_{n-2})$: each $c_i$ is a 
	        commitment of the secret share $s_i$.
		    All parties receive these values.
    \item $\mathrm{com}_{w}$: a commitment of the $P_{n-1}$'s 
          witness $w$. This value is revealed for all parties.
\end{itemize}

We make two changes in the second phase, i.e., the fair reconstruction protocol.
The first one is to add a condition for $P_{n-1}$ to refund the transactions created in
step 3), i.e., the transactions claimed in step 6). (See Protocol \ref{Prtcl:Ours}. The relevant transactions are \eqref{Eq:our-n+1} to \eqref{Eq:our-2n-2}.)
In our solution, $P_{n-1}$ must reveal the secret share $s_i$ within $\tau'_2 \,(>\tau_2)$ to refund the coins from $P_i$.
Note that the commitments $(c_1,\dots, c_{n-2})$ are used to verify the validity in this refund phase.
We do not change for the other transactions.

The second change is to add the following transactions.
(Suppose that $\tau''_2 > \tau'_2$.)
\begin{align}
	P_{n-1} \xrightarrow
	[\,\,\,\,\,\,\,\,\,\,\,\,\,\,\,(n-1)q, \tau''_2 \,\,\,\,\,\,\,\,\,\,\,\,\,\,\,]{w} P_1 
	\nonumber \\
	P_{n-1} \xrightarrow
	[\,\,\,\,\,\,\,\,\,\,\,\,\,\,\,(n-1)q, \tau''_2\,\,\,\,\,\,\,\,\,\,\,\,\,\,\,]{w} P_{2}  \nonumber \\
	\vdots \,\,\,\,\,\,\,\,\,\,\,\,\,\,\,\,\,\,\,\,\,\,\,\,\,\,\,\,\,\,\nonumber
	\\
	P_{n-1} \xrightarrow
	[\,\,\,\,\,\,\,\,\,\,\,\,\,\,\,(n-1)q, \tau''_2\,\,\,\,\,\,\,\,\,\,\,\,\,\,\,]{w} P_{n-2} \nonumber
\end{align}
These transactions are prepared to penalize $P_{n-1}$ when he/she refunds two or more
transactions in step 6).
Recall that $P_{n-1}$ must publish the secret share $S_i$ to refund the coins from $P_i$. 
Since the threshold value is two, the secret value $w$ is revealed for all parties if $P_{n-1}$ refunds two or more transactions in step 6).
Further, the middle parties can claim the added transactions when they learn $w$.
(Note that the commitment $\mathrm{com}_{w}$ is used here to verify the validities.)
Thus, the added transactions prevent $P_{n-1}$ from the excessive refund.

We confirm that the above-mentioned way obtains the equivalent compensations below.
If $P_{n-1}$ refunds $x$ transactions ($2 \leq x \leq n-2$) and gets back $\coins(x(n-1)q)$,
then he/she loses $\coins((n-2)(n-1)q)$, which is more than 
or equal to $\coins(x(n-1)q)$ for all $x$.
Namely, the added transactions prevent $P_{n-1}$ to refund two or more
transactions in step 6).
Also, if corrupted $P_{n-1}$ refunds two or more transactions in step 6), then the compensation amounts are $\coins(nq)$ for all honest parties.
Thus, the compensation amounts always are the same for all honest parties.
As a result, we resolve the non-equivalence and obtain secure computation with 
penalties in $O(1)$ rounds and $O(n)$ broadcasts.

We note that $P_{n-1}$ needs to create the added transactions before step 4), i.e.,
before transactions \eqref{Eq:our-2n-1} 
to \eqref{Eq:our-3n-4} are created.
Otherwise, corrupted $P_{n-1}$ can initiate step 5) without making the transactions.
Since it is sufficient to create the added transactions at the same round as steps 1)-3),
our solution can be applied without increasing the number of rounds. (We suppose that the added transactions are created in step 3) below.)

In summary, our fair reconstruction protocol that
resolves the non-equivalence proceeds as follows.\footnote{For a deposit transaction that does not specify a predicate for refund, we suppose that the predicate is set as a tautology.}

\medskip
\noindent
\textbf{Deposit phase:}
\begin{enumerate}
		\item For $i \in \{1, \dots, n-1\}$, $P_i$ makes a transaction to send $P_n$ $\coins(q)$
		with a predicate for claim $\phi_{i,n}$ and a round number for claim $\tau_4$,
		where $\phi_{i,n}(x) = 1$ only if $x = T_1 \wedge \dots \wedge T_n$.
		\item $P_n$ makes a transaction to send $P_{n-1}$ $\coins((n-1)q)$
		with a predicate for claim $\phi_{n,n-1}$ and a round number for claim $\tau_3$,
		where $\phi_{n,n-1}(x) = 1$ only if $x = T_1 \wedge \dots \wedge T_{n-1}$.
		\item For $i \in \{1,\dots,n-2\}$,
		$P_{n-1}$ makes a transaction to send $P_i$ $\coins((n-1)q)$
		with a predicate for claim $\phi_{n-1,i}$,
		a predicate for refund $\phi'_{n-1,i}$,
		a round number for claim $\tau_2$, and
		a round number for refund $\tau'_2$
		where $\phi_{n-1,i}(x) = 1$ only if $x = T_{n-1} \wedge T_i$ and $\phi'_{n-1,i}(x) = 1$ only if $x = S_i$.
		Further, $P_{n-1}$ makes a transaction to send $P_i$
		$\coins((n-1)q)$ with a predicate for claim $\psi_{n-1,i}$
		a round number for claim $\tau''_2$, where $\psi_{n-1,i}(x) = 1$ only if $x = w$.
		\item For $i \in \{1,\dots,n-2\}$, $P_i$ makes a transaction to send $P_{n-1}$
		$\coins((n-2)q)$
		with a circuit $\phi_{i,n-1}$ and a round number $\tau_1$,
		where $\phi_{i,n-1}(x) = 1$ only if $x = T_{n-1}$.
	\end{enumerate}
\medskip
\noindent
\textbf{Claim phase:}
	\begin{enumerate}	
	\setcounter{enumi}{4}
		\item $P_{n-1}$ claims by publishing $T_{n-1}$ in round $\tau_1$ and receives $\coins((n-2)q)$ 
		from each of $P_1, \dots, P_{n-2}$.
		\item For $i \in \{1,\dots,n-2\}$,
		$P_i$ claims by publishing $T_{n-1} \wedge T_i$ in round $\tau_2$ and receives 
		$\coins((n-1)q)$ from $P_{n-1}$.
		\item $P_{n-1}$ claims by publishing $T_1 \wedge \dots \wedge T_{n-1}$
		in round $\tau_3$ and receives $\coins((n-1)q)$ from $P_n$.
		\item $P_n$ claims by publishing $T_1 \wedge \dots \wedge T_n$ in round $\tau_4$
		and receives $\coins(q)$ from each of $P_1, \dots, P_{n-1}$.
\end{enumerate}
Note that there is no change in the claim phase from the protocol described in Section \ref{Sect:OurFair}
if all parties behave honestly.
In the claim phase, the procedure changes only when adversarial parties abort in step 6).
For the deposit phase, the change is in step 3) only. In step 3), $P_{n-1}$ creates two deposit transactions
for each of $P_1,\dots, P_{n-1}$ in one round.

\subsubsection*{Remark 3}
In the functionality of secure computation with penalties (Functionality \ref{Alg:Ff}), the compensation amount is always $q$. On the other hand, we allow the compensation amount to be $aq$ for $a \geq 1$, where $a$ is chosen by a simulator. We note that adversaries always choose $a = 1$ if they are rational since they want to minimize the compensation amount. Thus, if adversaries are rational, there is no difference between the original functionality and ours. We remark that the compensation amounts for honest parties are always the same regardless of the value $a$.

\section{Conclusion}\label{Sect:CR}
This paper focused on secure computation with penalties based on Bitcoin.
Bentov and Kumaresan \cite{Bentov14} showed that secure computation with penalties can be constructed with
$O(n)$ rounds and $O(n)$ broadcasts
for any function in the $(\mathcal{F}_{\mathrm{OT}}, \Fcr)$-hybrid model.
As far as we know, 
no protocol achieves $O(1)$ rounds and $O(n)$ broadcasts.

This paper showed a first protocol that needs only $O(1)$ rounds and $O(n)$ broadcasts.
First, we introduced secure computation with non-equivalent penalties that is a relaxed variant of secure computation with penalties in terms of the compensation amount.
In secure computation with penalties, every honest party can be compensated with the same amount of coins when an adversary aborts after learning the output value. On the other hand, in our setting, every honest party is guaranteed to be compensated with more than a predetermined amount of coins, but not the same amount. We showed that secure computation with non-equivalent penalties can be realized with $O(1)$ rounds and $O(n)$ broadcasts for arbitrary functions in the $(\mathcal{F}_{\mathrm{OT}}, \Fcr)$-hybrid model. In particular, we improved the round complexity of the fair reconstruction protocol, which is a key ingredient for realizing secure computation with penalties.

In addition to the above result, we showed two techniques to solve issues of our protocol.
First technique was to solve the issue of increasing the deposit amount to $O(n^2q)$ from $O(nq)$, which is the deposit amount of the Bentov-Kumaresan's protocol. 
Our technique reduces the deposit amount by about $1/(l+1)$ instead of the number of rounds is increased by $2l$.
We leave the following open problem: Is it possible to design secure computation with penalties that needs $O(1)$ rounds, $O(n)$ broadcasts, and $O(nq)$ deposit amount?

The second technique was to solve the non-equivalence of the compensation amount. In order to achieve this result, we proposed new functionality called claim-refund-or-give functionality. 
It achieved the equivalent compensation without sacrificing efficiency. Thus, a fair protocol can be realized with $O(1)$ rounds and $O(n)$ broadcasts based on Bitcoin.

\bibliography{bib.bib}

\appendix

\section{Proof of Theorem 1} \label{Sect:appendix}
In the SCC model, a simulator $\Sim$ needs to simulate two parts: one is the standard functionality for $f$ and 
another one is coins.

\subsection{Simulation for the standard functionality}
The simulation for the former is almost the same as the proof of Bentov-Kumaresan's protocol \cite{Bentov14ep}.
A simulator $\Sim$ performs one of the two simulations depending on whether $P_n$ is honest.
If $P_n$ is honest, then an adversary $\Adv$'s 
decision to abort
is independent of the output because of the assumption of {\sf pubNMSS}.
In this case, $\Sim$ can simulate in the same way as in standard secure computation.
If $P_n$ is corrupted, then $\Adv$'s decision to abort
can depend on the output. In order to simulate this, $\Sim$ must send $\mFf$
corrupted parties' inputs to receive the output. 
Although the output and the shares distributed to corrupted parties may contradict,
$\Sim$ can simulate it by using the equivocation
property of the honest-binding commitment.

\subsection{Simulation for coins}
We here show how to simulate coins.
In the SCC model, an environment $\Env$ can initialize parties' wallets and choose parties' inputs. 
We note that $\Sim$ cannot create coins, 
and thus
it must complete the simulation using only the coins provided by $\Env$.
Suppose that $\Env$ provides each party with the minimum amount of coins necessary to run the protocol. If $\Env$ provides fewer coins than this, then $\Sim$ terminates the simulation.
We use $\SW$, which accounts for coins provided by corrupted parties. 
At the beginning of the simulation,  
$\SW$ has no coins since all coins are provided to $\Adv$.
$\SW$ must hold a compensation amount 
to run $\mFf$ when an adversary aborts to steal the output,
i.e., aborts without telling the value to honest parties.
In secure computation with penalties, 
when $\Adv$ steals the output, $\Sim$ 
sends $\coins(h'q)$ to $\Ff$ to learn the output.
Note that $h' = |H'|$ means the number of honest parties who should be compensated.
On the other hand, in secure computation with
non-equivalent penalties, when $\Adv$ aborts, $\Sim$ sends $\mFf$
\textit{more than} $\coins(h'q)$, which is the 
main difference from the proof of Bentov-Kumaresan's protocol.

If an adversary $\Adv$ aborts in the deposit phase,
a simulator $\Sim$ sends all coins in $\SW$ to $\Adv$. 
And then it terminates the simulation.
Next, suppose that the adversary performs honestly
in the deposit phase.
Note that $\SW$ has all coins deposited by $\Adv$
at the beginning of the claim phase.

Let us consider the case where an adversary $\Adv$ aborts
in the claim phase.
As mentioned above, $P_{n-1}$ is the only party to 
be able to receive more than $\coins(q)$ as compensation.
Also, the deposit amount is $\coins((n^2-3n+3)q)$ for $P_{n-1}$
and $\coins((n-1)q)$ for all other parties.
Hence, we discuss the claim phase separating the cases where $n-1 \in C$ or not.
Also, we suppose that $n \in C$ since we mainly discuss
the case where $\Adv$ aborts to steal the output.

\medskip
\noindent
\textbf{Case of $n-1 \in C$:}
In this case, all honest parties in $H'$ must
receive $\coins(q)$ as compensation.
Note that all of them belong to the middle parties. 
At the beginning of the claim phase,
$\SW$ holds $\coins((c-1)(n-1)q+(n^2-3n+3)q)$ 
that are the
sum of deposited coins by the corrupted parties, where $c=|C|$.
If corrupted $P_{n-1}$ reveals $T_{n-1}$ honestly, then $\Sim$ sends
$\coins((n-2)^2q)$, which are taken from $\SW$, to $\Adv$.
As a result, $\SW$ holds $\coins(c(n-1)q)$.
Let us consider the case where $\Adv$ aborts in step 6) (see Section \ref{Sect:OurFair}) and steals the output.
 For the sake of simplicity, we here assume that every corrupted party aborts in step 6). \footnote{Note that $\Adv$ only needs one of the corrupted parties aborts in step 6) to steal the output. $\Sim$ can also similarly simulate such a case.}
Then, $\Sim$ sends $\coins((n-c)q)$ to $\mFf$ to learn the output.
Furthermore, $\Sim$ here needs to send $\Adv$ coins to simulate refunds for corrupted parties.
The transactions that are not claimed consist of
(A) ones created by corrupted parties in steps 1)--2) 
and (B) ones created in step 3) by $P_{n-1}$ 
to corrupted parties.
The amount of coins in (A) and (B) are
$\coins((c-1)q+(n-1)q)$ and $\coins((n-1)(c-2)q)$, respectively.
Namely, $\Sim$ sends $\coins(n(c-1)q)$,
which is the sum of (A) and (B), to $\Adv$.
Since $\Sim$ must send $\coins((n-c)q)$ to $\mFf$
and $\coins(n(c-1)q)$ to $\Adv$,
$\SW$ must hold $\coins(c(n-1)q)$.
This is the same amount of coins $\SW$ holds.
Thus, it completes to simulate the case
where $\Adv$ aborts to steal the output in step 6).
    
If an adversary $\Adv$ behaves honestly in
step 6), then a simulator $\Sim$ sends $\coins((n-1)(c-2)q)$ to $\Adv$.
 Also, suppose that $\Adv$ behaves in step 7) honestly. Then, $\Sim$ sends $\coins((n-1)q)$
to $\Adv$.\footnote{Since $P_n$ and $P_{n-1}$
are corrupted, $\Sim$ sends the same coins to $\Adv$
as refunded if $P_{n-1}$ aborts in this step.}
As a result, $\SW$ holds $\coins((n-1)q)$.
Let us consider the case where $\Adv$ aborts in step 8) and steals the output.
Then, $\Sim$ sends $\mFf$ $\coins((n-c)q)$ to learn the output. Furthermore, $\Sim$ sends $\coins((c-1)q)$, which is the refund, to $\Adv$.
Thus, $\SW$ must hold $\coins((n-1)q)=\coins((n-c)q+(c-1)q)$.
This is the same amount of coins $\SW$ holds.
Thus, it completes to simulate the case
where $\Adv$ aborts to steal the output in step 8).    
    
\medskip
\noindent
\textbf{Case of $n-1 \notin C$:}
At the beginning of the claim phase,
$\SW$ holds $\coins(c(n-1)q)$.
Note that $\Sim$ uses no coins in $\SW$ to simulate step (5)
since $P_{n-1}$ is honest.
Let us consider the case where an adversary $\Adv$ 
aborts in step 6) and steals the output.
$P_{n-1}$ is compensated with 
$\coins((nc-2n-c+3)q) = \coins((n-2)^2q-(n-c-1)(n-1)q)$.
The other honest parties are compensated with
$\coins(q)$.
Thus, $\Sim$ sends $\coins((nc-n-2c+2)q)$ to $\mFf$ to learn the output. Furthermore, $\Sim$ sends $\coins((n+c-2)q)$, which is the refund, to $\Adv$.
$\SW$ must hold $\coins(c(n-1)q) = \coins((nc-n-2c+2)q) + \coins((n+c-2)q)$, and this is the same amount of coins $\SW$ holds.
Thus, it completes to simulate the case
where $\Adv$ aborts to steal the output in step 6).

Suppose that an adversary behaves honestly in steps 6) and 7).
Then, $\Sim$ sends $\coins((nc-n-c+1)q)$ to $\Adv$ to simulate step 6) and no coins to simulate step 7).
As a result, $\SW$ holds $\coins((n-1)q)$.
Let us consider the case where $\Adv$ aborts in step 8) and steals the output.
Then, $\Sim$ sends $\mFf$ $\coins((n-c)q)$ to learn the output. Furthermore, $\Sim$ sends $\coins((c-1)q)$, which is the refund, to $\Adv$.
Thus, $\SW$ must hold $\coins((n-1)q)=\coins((n-c)q+(c-1)q)$.
This is the same amount of coins $\SW$ holds.
Thus, it completes to simulate the case
where $\Adv$ aborts to steal the output in step 8).  
\hfill $\Box$

\section{Realization of $\Fcrg$ via Bitcoin}
Suppose that a Bitcoin transaction consists of $(id,x,\sigma,\pi,\tau)$, where 
$id$ is an identifier of a previous transaction, $x$ is the amount of coins, $\sigma$ is an input script,
$\pi$ is an output script, and $\tau$ is a time-bound.
For a Bitcoin transaction $txn$, $\mathrm{simp}(txn)$ means a \textit{simplified form} of $txn$.
The simplified form refers to $(id,x,\pi,\tau)$, i.e., it expresses a transaction excluding its input script.
Note that Bitcoin script opcodes are allowed to take a simplified form of a transaction itself as an input.

Protocol \ref{Prtcl:CRG1} shows an implementation of the claim-refund-or-give functionality $\Fcrg$.
In this implementation, the parties create four transactions, $txn_{\mathrm{CRG}}$,
$txn_{\mathrm{claim}}$, $txn_{\mathrm{refund}}$, $txn_{\mathrm{give}}$.
$P_s$ broadcasts transaction $txn_{\mathrm{CRG}}$ in the deposit phase.
The remaining three transactions are used to redeem $txn_{\mathrm{CRG}}$.
Depending on which of the claim, refund, or give is executed, one of the 
three transactions is broadcast to the Bitcoin network.

See the opcode of $txn_{\mathrm{CRG}}$ in step 4).
We can read that this opcode consists of three parts by dividing the opcode by OR operation:
\begin{itemize}
	\item $\mathrm{OP\_CHECKSIG}(pk_r,\cdot)$ AND $\phisr(\cdot)$
	\item $\mathrm{OP\_CHECKSIG}(pk_r,\cdot)$ AND $\mathrm{OP\_CHECKSIG}(pk_s,\cdot)$ AND $\phirf(\cdot)$
	\item $\mathrm{OP\_CHECKSIG}(pk_r,\cdot)$ AND $\mathrm{OP\_CHECKSIG}(pk_s,\cdot)$ AND $H(\cdot)$
\end{itemize}
Thus, a party can redeem $txn_{\mathrm{CRG}}$ only if he/she provides an input script that satisfies
one of the three conditions.

The first opcode is prepared for the claim phase. In order to redeem $txn_{\mathrm{CRG}}$ using
this opcode, $P_r$ must reveals a witness $w_r$ such as $\phisr(w_r)=1$.

The second opcode is prepared for the refund phase.
In order to redeem $txn_{\mathrm{CRG}}$ using
this opcode, $P_s$ must reveal a witness $w_s$ such as $\phirf(w_s)=1$.
Also, $P_s$ uses $sig_r$, which is generated by $P_r$ in step 12), to
clear the opcode $\mathrm{OP\_CHECKSIG}(pk_s,\cdot)$.
$\mathrm{OP\_CHECKSIG}(pk_r,\cdot)$ is prepared to prevent $P_s$ from refunding before $\tau_{\mathrm{claim}}$.

The third opcode is for the give phase.
If $txn_{\mathrm{CRG}}$ is not redeemed at $\tau_{\mathrm{refund}} + 1$,
then $P_r$ can receive the coins by revealing the random value $\lambda$ that is generated by 
$P_r$ in step 2).
$P_r$ uses $sig_s$, which is generated by $P_s$ in step 8), to
satisfy the opcode $\mathrm{OP\_CHECKSIG}(pk_s,\cdot)$.
$\mathrm{OP\_CHECKSIG}(pk_s,\cdot)$ is used to prevent $P_r$ from refunding before $\tau_{\mathrm{refund}}$ by redeeming $txn_{\mathrm{CRG}}$ with other than $txn_{\mathrm{give}}$.
Also, $H(\cdot)$ prevents $P_s$ from redeeming $txn_{\mathrm{CRG}}$
by using the third opcode.
If $H(\cdot)$ is not included the third opcode,
$P_s$ can redeem $txn_{\mathrm{CRG}}$ by using $sig_r$ in step 12).

\begin{algorithm}[ht]
	\floatname{algorithm}{Protocol}
	\caption{Implementation of $\Fcrg$ via Bitcoin.}
	\label{Prtcl:CRG1}		
\begin{description}
    \item[\textbf{Deposit phase}:]
    Perform the following process.
	\begin{enumerate}
		\item $P_s$ requests a fresh public key by sending
			  $(\mathsf{deposit\_init},sid,ssid,s,r,\tau_{\mathrm{claim}},\tau_{\mathrm{refund}})$
			  to $P_r$.
		\item $P_r$ generates a fresh $(sk_r,pk_r)$ pair, a random value 
			  $\lambda \in \{0,1\}^k$, and a predicate $\phirf$.
		\item $P_r$ sends
		       $(\mathsf{deposit\_ack},sid,ssid,s,r,\tau_{\mathrm{claim}},\tau_{\mathrm{refund}},pk_r,
		       H(\lambda),\phirf)$ to $P_s$ where $H$ is a hash function.
		\item $P_s$ takes his/her public key $pk_s$ and creates a Bitcoin transaction 
		      $txn_{\mathrm{CRG}}$ that redeems $\coins(c)$. The output script $\pi$ is set as
		      follows:
		      \vspace{-1mm}
		      \begin{align*}
		      	&\pi(\cdot) = \mathsf{OP\_CHECKSIG}(pk_r,\cdot) 
		      	\,\,\mathrm{AND}\,\,\\
		      	& \,\,\,\,\,\,\,\,\,\,\,\,\,\,\,\,
		      	(\phisr(\cdot)  \,\, \mathrm{OR}\,\,
		      	(\mathrm{OP\_CHECKSIG}(pk_s,\cdot) \,\,\mathrm{AND}\,\,\\
		      	& \,\,\,\,\,\,\,\,\,\,\,\,\,\,\,\,\,\,\phirf(\cdot))
		      	\,\, \mathrm{OR}\,\,
		      	(\mathrm{OP\_CHECKSIG}(pk_s,\cdot) \,\,\mathrm{AND}\,\, H(\cdot)))
		      \end{align*}
		      \vspace{-5mm}
	      \item $P_s$ computes the identifier $id_{\mathrm{CRG}}$ of $txn_{\mathrm{CRG}}$
	      		and sends it to $P_r$.
	      \item $P_r$ prepares a transaction $txn_{\mathrm{give}}$ that takes 
	      		$id_{\mathrm{CRG}}$ as its input script.
	     		$txn_{\mathrm{give}}$ has the locktime $\tau_{\mathrm{refund}}$
	     		and an output script $\pi'(\cdot)$ that $P_r$ controls, i.e.,
	     		$\mathrm{simp}(txn_{\mathrm{give}}) := (id_{\mathrm{CRG}},x,\pi',\tau_{\mathrm{refund}})$.
	      \item $P_r$ sends			
	       	$(\mathsf{deposit\_sign\_give},sid,ssid,s,r,\mathrm{simp}(txn_{\mathrm{give}}))$
	      		to $P_s$.	
	      \item $P_s$ computes	 
	     		$sig_s := \mathrm{Sign}_{sk_s}(\mathrm{simp}(txn_{\mathrm{give}}))$.
	      		and sends $(\mathsf{deposit\_sign\_give\_ack},sid,ssid,s,r,sig_s)$ to $P_r$. 			
	      \item $P_r$ checks whether 
	            $\mathrm{Vrfy}_{pk_s}(\mathrm{simp}(txn_{\mathrm{give}}),sig_s) =1$.	    
	      \item $P_s$ prepares a transaction $txn_{\mathrm{refund}}$ that takes 
	     	 	$id_{\mathrm{CRG}}$ as its input script.
	      		$txn_{\mathrm{refund}}$ has the locktime $\tau_{\mathrm{claim}}$
	      		and an output script $\pi''(\cdot)$ that $P_s$ controls, i.e.,
	      		$\mathrm{simp}(txn_{\mathrm{refund}}) := (id_{\mathrm{CRG}},x,\pi'',\tau_{\mathrm{refund}})$.	    
	      \item $P_s$ sends			
	      		$(\mathsf{deposit\_sign\_refund},sid,ssid,s,r,\mathrm{simp}(txn_{\mathrm{refund}}))$
	      		to $P_r$.		    
	      \item	$P_r$ computes $sig_r := \mathrm{Sign}_{sk_r}(\mathrm{simp}(txn_{\mathrm{refund}}))$
	      		and sends $(\mathsf{deposit\_sign\_refund\_ack},sid,ssid,s,r,sig_r)$ to $P_s$. 
	      \item If $\mathrm{Vrfy}_{pk_r}(\mathrm{simp}(txn_{\mathrm{refund}}),sig_r) =1$,
	      		then $P_s$ broadcasts $txn_{\mathrm{CRG}}$ to the Bitcoin network.
	\end{enumerate}
	\vspace{-4mm}
	\item \textbf{Claim phase:}
		 $P_r$ broadcasts to the Bitcoin network a transaction $txn_{\mathrm{claim}}$
			  that redeems $txn_{\mathrm{CRG}}$, by providing 
		      $\mathrm{Sign}_{sk_r}(\mathrm{simp}(txn_{\mathrm{refund}}))$ and $w_r$ 
		      such as $\phisr(w_r)=1$.
	\vspace{-2mm}		      
	\item \textbf{Refund phase:}
		At $\tau_{\mathrm{claim}} + 1$, $P_s$ broadcasts to the Bitcoin network the transaction 
			  $txn_{\mathrm{refund}}$ that
		      redeems $txn_{\mathrm{CRG}}$, by providing $sig_r$, 
		      $\mathrm{Sign}_{sk_s}(\mathrm{simp}(txn_{\mathrm{refund}}))$, and
		      $w_s$ such as $\phirf(w_s)=1$. 	
	\vspace{-2mm}	      
	\item \textbf{Give phase:}
		At $\tau_{\mathrm{refund}} + 1$, $P_r$ broadcasts to the Bitcoin network the transaction 
			  $txn_{\mathrm{give}}$ that redeems $txn_{\mathrm{CRG}}$, by providing 
			  $\mathrm{Sign}_{sk_r}(\mathrm{simp}(txn_{\mathrm{give}}))$, $sig_s$, and $\lambda$. 
\end{description}
\end{algorithm}	



\end{document}